\documentclass[fleqn,usenatbib]{mnras}

\usepackage{newtxtext,newtxmath}

\usepackage[T1]{fontenc}

\DeclareRobustCommand{\VAN}[3]{#2}
\let\VANthebibliography\thebibliography
\def\thebibliography{\DeclareRobustCommand{\VAN}[3]{##3}\VANthebibliography}

\usepackage{graphicx}	
\usepackage{amsmath}	

\newcommand\nn         {\nonumber}

\def\mnras{{Mon.~ Not.~ R.~ Astron.~ Soc.~}}

\def\apj{{Astrophys.~ J.~}}
\def\apjs{{Astrophys.~ J.~ Suppl.~}}

\def\mnras{{MNRAS}}

\def\apj{{ApJ}}
\def\apjs{{ApJS}}

\def\aap{{A\&A}}

\def\physrep{{Phys.~ Rep.~}}
\def\jcap{Journal of Cosmology and Astro-Particle Physics}

\newcommand{\be}{\begin{equation}}
\newcommand{\ee}{\end{equation}}
\newcommand{\ba}{\begin{eqnarray}}
\newcommand{\ea}{\end{eqnarray}}

\def\pp1{{\prime}}
\def\pp2{{\prime\prime}}

\def\2D{{\rm 2D}}

\def\bk{{\bf k}}

\def\1Loop{{\rm 1Loop}}

\def\rhob{\bar{\rho}}

\def\kMpc{\, h \, {\rm Mpc}^{-1}}

\font\BF=cmmib10

\def\v{{\hbox{\BF v}}}

\def\fun#1#2{\lower3.6pt\vbox{\baselineskip0pt\lineskip.9pt
        \ialign{$\mathsurround=0pt#1\hfill##\hfil$\crcr#2\crcr\sim\crcr}}}

\newcommand{\bea}{\begin{eqnarray*}}
\newcommand{\eea}{\end{eqnarray*}}

\title[Universal mass function beyond $\Lambda$CDM]
      {On the universality of the halo mass function beyond $\Lambda$CDM cosmology}

\author[Y. Li \& R~E.~Smith]{
Yuhao Li$^{1}$\thanks{E-mail: yl700@sussex.ac.uk}
and
Robert E. Smith$^{1}$\thanks{E-mail: r.e.smith@sussex.ac.uk}
\\
$^{1}$Astronomy Centre, Department of Physics and Astronomy,
University of Sussex, Brighton, BN1 9RH, UK.  }

\date{Accepted 2025 June 12. Received 2025 May 30; in original form 2024 December 19}

\pubyear{2025}

\begin{document}
\label{firstpage}
\pagerange{\pageref{firstpage}--\pageref{lastpage}}
\maketitle


\begin{abstract}
  {Theoretical frameworks based on Press-Schechter formalism and excursion set arguments suggest that the abundance of dark matter haloes exhibits universal behaviour when expressed in terms of peak height. If true, this implies that a single high-accuracy cosmological simulation could serve as a basis for constructing an emulator applicable to any other cosmology of interest. This tantalising possibility has inspired numerous studies over the years. However, in practice, different ways of defining haloes have led to mixed results concerning this issue. In this work, we utilise a suite of high-resolution cosmological $N$-body simulations, to revisit this question for friends-of-friends haloes under the flat time-evolving $w$CDM model, with simple modifications of the primordial physics via variations in the scalar spectral index and its running. We construct a reference locus of $\nu f(\nu)$ from our fiducial $\Lambda$CDM simulation and compare it against measurements from alternative models. We find that deviations from the locus remain within $5\%$ when varying each of the parameters within the ranges: ${w_0} = -1.0 \pm 0.1$, $w_a = 0\pm0.2$, $\Omega_{\rm DE} = 0.693\pm0.050$, $\omega_{\rm c} = 0.119\pm0.006$, $\omega_{\rm b} =0.0222\pm0.0011$, $A_{\rm s} = (2.15\pm0.22) \times {10^{-9}}$, $n_{\rm s} =0.961\pm0.048$, $\alpha_{\rm s}\ = 0\pm0.01$, for redshift $z < 7$.}

\end{abstract}





\section{Introduction}\label{sec:intro}

One of the keystones of modern cosmology is that the majority of
gravitating mass in the Universe, some $\sim 85\%$, is in the form of
collisionless cold dark matter (hereafter CDM) with the remainder
being in the form of the known standard model particles `baryons' and
neutrinos \citep{Planck2014XVI}. The true physical nature of the dark
matter particle remains a mystery \citep[for a review
  see][]{Bertoneetal2010}. Primordial fluctuations in the CDM are
thought to be sourced after a period of inflation, whereupon they are
well described by a Gaussian random field with a power spectrum that
has almost scale-invariant potential fluctuations. {However, during the radiation-dominated era, fluctuations inside the horizon are suppressed due to the presence of radiation pressure, shaping the matter power spectrum we observe today. This suppression results in a characteristic turnover in the power spectrum, with a spectral slope approaching -3 on small scales rather than 1.} The wealth of
late-time cosmic structures that we observe today, the sheets,
filaments, and clusters -- the cosmic web -- then forms subsequently
through the gravitational instability of these processed primordial density seeds.

One of the commonly used theoretical frameworks to describe the
late-time distribution of matter is the halo model. This simply
divides all of the matter into a set of distinct units -- haloes
\citep[see,][for reviews]{CooraySheth2002,Asgarietal2023}. If one
considers the halo model in its {basic} form, where all particles are
considered to belong to a halo, going down to single particle haloes,
then it is simply an equivalence class on the matter. However, to
elevate this from a tautology and so make useful predictions one
requires a model for the abundance of haloes, their evolution with
time, their clustering, and their internal density distributions. With
these in hand one can make predictions for an astonishingly wide array
of cosmological observables \citep[for examples of applications
  see][]{CooraySheth2002,Asgarietal2023}. The halo model framework has
been of great utility for many of the wide field large-scale structure
surveys (especially if one encompasses SHAMs and HAMs etc). For
example, recently in the Dark Energy Survey it was extensively used in
\citet{Friedrichetal2021} for building covariance models, and for the
KiDS survey by \citet{Dvorniketal2023} who explored the galaxy matter
connection, among others. It will no doubt also play an important role
in the exploitation of the upcoming surveys like DESI
\citep[][]{DESI2016}, Euclid \citep[][]{Euclid2011} and LSST
\citep[][]{LSST2019} surveys. Thus to fully exploit the data from
these missions, there is high demand for theoretical models that can
accurately predict the abundance of dark matter haloes over a wide
range of masses and cosmic time scales to high precision. And doing so
for a wide array of viable scenarios beyond the vanilla flavoured
vacuum energy dominated CDM model (hereafter $\Lambda$CDM).
Exploring this is one of the aims of this paper.

An analytic form of the halo mass function (HMF) was first derived by
\citet{PressSchechter1974} by means of applying the spherical collapse
model to arbitrary points in the initial linear matter density
field. They noted that their model predicted that only half of the
mass in the Universe was in collapsed haloes, which they corrected in
an {{\em ad hoc}} fashion by multiplication of their final result by 2! Later
on, \citet{Bondetal1991} provided a robust derivation of their
formula, which included the missing factor, using excursion set
theory. Interestingly, it was found that the Press-Schechter HMF could
be rearranged into a universal form, i.e., being completely independent of cosmology and
redshift, through a judicious choice of variables. Although the
Press-Schechter function failed to accurately predict the number
densities of very large and small mass haloes that were measured in
cosmological simulations
\citep[][]{LaceyCole1994,ShethTormen1999,Jenkinsetal2001}, the
possibility of a universal HMF inspired a number of subsequent
studies.

The path to a universal HMF is barred by a number of hurdles, one of
which is agreeing what the actual definition of a halo is and its
associated mass \citep[see][for a discussion]{White2002} \citep[and
  for a review of methods and codes see][]{Knebeetal2011}:
\begin{itemize}
  \item {\bf FoF haloes}: One of the earliest operational definitions
    was given through the Friends-of-Friends (FoF) algorithm of
    \citet{Davisetal1985}. This identifies haloes in numerical
    simulations as all linked particles with separations less than
    $b\bar{n}^{-1/3}$, where $\bar{n}$ is the mean particle density
    and $b$ is the linking parameter typically set to $b=0.2$.  This
    halo finder makes no assumptions about the shape of the halo and
    is computationally efficient and fast. {The method directly groups particles without needing to smooth densities, estimate continuous fields, or iteratively determine the centre and boundary of the halo,} but can suffer from `bridging'
    between neighbouring groups. Subsequent studies of FoF haloes show
    that the mass functions of these objects can exhibit, approximate,
    universal behaviour
    \citep{ShethTormen1999,Jenkinsetal2001,White2002,Reedetal2003,
      Warrenetal2006,Reedetal2007,Lukicetal2007, Crocceetal2010,
      Maneraetal2010, Courtinetal2011, Bhattacharyaetal2011,
      Juanetal2014}.

\item {\bf SO haloes}: Another common halo finder, that is possibly
  more physically motivated, is the spherical overdensity (SO) method,
  which identifies haloes as all matter within a sphere whose average
  density is some fraction $\Delta$ of the background density
  \citep{LaceyCole1994}. This was inspired by the spherical collapse
  model, for which, in the Einstein de-Sitter model, after
  virialisation the average overdensity is found to be
  $\sim180\rho_{\rm crit}$, where $ \rho_{\rm crit}$ is the critical
  density. In many studies the characteristic overdensity is typically
  fixed at $\Delta=200$ for all cosmological model, this is in
  contrast to the spherical collapse density which varies with the
  underlying model \citep{LaceyCole1993,Ekeetal1996}. {On the other hand, in practice the algorithms using SO approach to define halo boundaries may differ in identifying the centre of halo \citep[see,][for a review]{EuclidXXIV2023}. Subsequent
  numerical studies of SO haloes show that with the exception of virial overdensity, the mass functions of these
  haloes more likely departs from universal behaviour \citep{Tinkeretal2008,
    Watsonetal2013, Despalietal2015, McClintocketal2019,Ondaro-Malleaetal2021,Gavas:2022iqb} in Einstein-de Sitter and $\Lambda$CDM cosmologies. In contrast, a different trend was observed in modified gravity scenarios \citep{Gupta:2021pdy}}.

\item {{\bf Other haloes}: More recently, another two, perhaps even more    physically-motivated, models for halo finding were proposed. One involves using     the caustic splashback radius   \citep{DiemerKravtsov2014,Adhikarietal2014,Moreetal2015}. The splashback
  radius is located at the place where the accreted matter reaches its
  first orbital apocentre after turnaround, which corresponds to a
  sharp drop in the halo density profile -- in the case of perfect spherical symmetry forming a zero density caustic. All matter within this
  radius is then considered to be part of the halo. The other one defines halo as a collection of particles orbiting in their own self-generated potential \citep{Garcia:2022zsz}. A subsequent
  study, has shown that haloes defined using the splashback-radius
  method are also potentially good candidates in favour of
  universality \citep{Diemeretal2020}. }

\end{itemize}

In summary, these studies have shown that haloes identified through
the FoF method, are a good candidate for exploring the universality of
the HMF. They also have the practical advantage over the splashback
haloes of having a number of computationally efficient algorithms for
measuring them in numerical simulations.  In this work, we revisit the
universality of the HMF with respect to FoF haloes
(with $b=0.2$) in the $\Lambda$CDM framework. We also explore a
variety of extensions beyond this framework, including time evolving
dark energy, and modifications of the primordial physics including
running of the primordial power spectral index. We do this by making use
of the suite of high-resolution cosmological $N$-body simulations, the
D\"ammerung Simulations, carried out by
\citet{SmithAngulo2018}. Unlike previous relevant studies, we do not
assume any existing functional form of the universal HMF. But instead
we do assume that a universal locus of the HMF exists, which in
practice we take from our fiducial $\Lambda$CDM suite of runs and
compare this with results from our variations in the underlying
cosmological model.

This paper is organised as follows: In Section \S\ref{sec:sims} we
describe the D\"ammerung simulation suite used in this work. In
\S\ref{sec:hmf} we briefly introduce the theoretical background,
describe our estimators for the mass function and their associated
uncertainties, and how we test the universality with respect to
cosmology and redshift. In \S\ref{sec:results} we present our results
for the set of models and redshifts that we consider in our study, and
discuss their implications. Finally, in \S\ref{sec:conclusions} we
conclude our findings and discuss future directions.


\begin{table*}
	\centering
	\caption{Cosmological parameters used in the
          D\"ammerung simulations. Column 1 denotes the simulation
          series. The next 8 columns denote the 8 cosmological
          parameters listed in Eq.~(\ref{eq:2.1}) that define each simulated
          model: columns (2) and (3) give $w_0$ and $w_a$ which are
          parameters used to define the dark energy equation of state;
          column (4) gives the density parameter for dark energy
          $\Omega_{\rm DE}$; columns (5) and (6) give the physical
          densities of CDM $\omega_{\rm c}=\Omega_{\rm c} h^2$ and
          baryons $\omega_{\rm b} = \Omega_{\rm b}h^2$; columns (7), (8) and
          (9) give the primordial amplitude of scalar perturbations
          $A_{\rm s}$, the spectral index $n_{\rm s}$ and the running
          of the spectral index $\alpha_{\rm s}$. Columns (10)-(12) 
          give $\Omega_{\rm m}$, $h$ and $\sigma_{8}$, all of which 
          are derived parameters.  
          Parameter values that are deviations from
          the fiducial model are highlighted in bold.}
	\label{tab:1}
	\begin{tabular}{ccccccccc @{\ \ \ \vline\ \ \  } ccc}
		\hline
		Parameter & $w_{0}$ & $w_{a}$ & $\Omega_{\rm DE}$ & $\omega_{\rm c}$ & $\omega_{\rm b}$ & $A_{\rm s} [\times 10^{-9}]$ & $n_{\rm s}$ & $\alpha_{\rm s}$ & $\Omega_{\rm m}$ & $h$ & $\sigma_{8}$ \\
		\hline
		\hline
		Fiducial & $-1.0$ & $0.0$ & $0.6929$ & $0.1189$ & $0.02216$ & $2.148$ & $0.9611$ & $0.0$ & $0.3071$ & $0.6777$ & $0.8284$ \\
		\hline
		V1 & $\mathbf{-1.1}$ & $0.0$ & $0.6929$ & $0.1189$ & $0.02216$ & $2.148$ & $0.9611$ & $0.0$ & $0.3071$ & $0.6777$ & {\bf 0.8477} \\
		V2 & $\mathbf{-0.9}$ & $0.0$ & $0.6929$ & $0.1189$ & $0.02216$ & $2.148$ & $0.9611$ & $0.0$ & $0.3071$ & $0.6777$ & {\bf 0.8052} \\
		\hline
		V3 & $-1.0$ & $\mathbf{0.2}$ & $0.6929$ & $0.1189$ & $0.02216$ & $2.148$ & $0.9611$ & $0.0$ & $0.3071$ & $0.6777$ & {\bf 0.8136} \\
		V4 & $-1.0$ & $\mathbf{-0.2}$ & $0.6929$ & $0.1189$ & $0.02216$ & $2.148$ & $0.9611$ & $0.0$ & $0.3071$ & $0.6777$ & {\bf 0.8400} \\
		\hline
		V5 & $-1.0$ & $0.0$ & $\mathbf{0.7429}$ & $0.1189$ & $0.02216$ & $2.148$ & $0.9611$ & $0.0$ & ${\bf 0.2571}$ & $\mathbf{0.7407}$ & {\bf 0.8438} \\
		V6 & $-1.0$ & $0.0$ & $\mathbf{0.6429}$ & $0.1189$ & $0.02216$ & $2.148$ & $0.9611$ & $0.0$ & ${\bf 0.3571}$ & $\mathbf{0.6285}$ & {\bf 0.8137} \\
		\hline
		V7 & $-1.0$ & $0.0$ & $0.6929$ & $\mathbf{0.1248}$ & $0.02216$ & $2.148$ & $0.9611$ & $0.0$ & $0.3071$ & $\mathbf{0.6918}$ & {\bf 0.8580} \\
		V8 & $-1.0$ & $0.0$ & $0.6929$ & $\mathbf{0.1129}$ & $0.02216$ & $2.148$ & $0.9611$ & $0.0$ & $0.3071$ & $\mathbf{0.6633}$ & {\bf 0.7981} \\		
		\hline
		V9 & $-1.0$ & $0.0$ & $0.6929$ & $0.1189$ & $\mathbf{0.02327}$ & $2.148$ & $0.9611$ & $0.0$ & $0.3071$ & $\mathbf{0.6804}$ & {\bf 0.8216} \\
		V10 & $-1.0$ & $0.0$ & $0.6929$ & $0.1189$ & $\mathbf{0.02105}$ & $2.148$ & $0.9611$ & $0.0$ & $0.3071$ & $\mathbf{0.6750}$ & {\bf 0.8352} \\		
		\hline
		V11 & $-1.0$ & $0.0$ & $0.6929$ & $0.1189$ & $0.02216$ & $\mathbf{2.363}$ & $0.9611$ & $0.0$ & $0.3071$ & $0.6777$ & {\bf 0.8688} \\
		V12 & $-1.0$ & $0.0$ & $0.6929$ & $0.1189$ & $0.02216$ & $\mathbf{1.933}$ & $0.9611$ & $0.0$ & $0.3071$ & $0.6777$ & {\bf 0.7858} \\		
		\hline
		V13 & $-1.0$ & $0.0$ & $0.6929$ & $0.1189$ & $0.02216$ & $2.148$ & $\mathbf{1.0092}$ & $0.0$ & $0.3071$ & $0.6777$ & {\bf 0.8438} \\
		V14 & $-1.0$ & $0.0$ & $0.6929$ & $0.1189$ & $0.02216$ & $2.148$ & $\mathbf{0.9130}$ & $0.0$ & $0.3071$ & $0.6777$ & {\bf 0.8136} \\		
		\hline
		V15 & $-1.0$ & $0.0$ & $0.6929$ & $0.1189$ & $0.02216$ & $2.148$ & $0.9611$ & $\mathbf{0.01}$ & $0.3071$ & $0.6777$ & {\bf 0.8304} \\
		V16 & $-1.0$ & $0.0$ & $0.6929$ & $0.1189$ & $0.02216$ & $2.148$ & $0.9611$ & $\mathbf{-0.01}$ & $0.3071$ & $0.6777$ & {\bf 0.8263} \\
		\hline
	\end{tabular}
\end{table*}


\section{Numerical simulations}\label{sec:sims}

We start with introducing the cosmological model of the D\"ammerung
simulations, followed by the specifications of the simulation
settings. D\"ammerung simulations are a suite of dark-matter-only (DMO)
cosmological $N$-body simulations carried out by
\citet{SmithAngulo2018}, which assume a flat, dark energy dominated cold
dark matter model (hereafter $w$CDM), characterised by 8 parameters as
follows:
\begin{equation} 
p_{\alpha} = \{w_0, w_a, \Omega_{\rm DE}, \omega_{\rm c},
\omega_{\rm b}, A_{\rm s}, n_{\rm s}, \alpha_{\rm s}\}\ \ ,
    \label{eq:2.1}
\end{equation} 
where $w_0$ and $w_a$ define the dark energy equation of state parameter $w(a)$ via 
\begin{equation}
w(a) = w_0 + w_a (1 - a), \label{eq:2.2}
\end{equation} 
as proposed by \citet{ChevallierPolarski2001}.  $\Omega_{\rm DE}$ is
the present day density parameter for dark energy, $\omega_{\rm c} =
\Omega_{\rm c} h^2$ and $\omega_{\rm b} = \Omega_{\rm b} h^2$ are the
physical densities of cold dark matter and baryons, and $h$ is the
dimensionless Hubble parameter.

The matter power spectrum is initialised by specifying the primordial
power spectrum of curvature perturbations and we make use of the
following form \citep{Komatsuetal2009,Planck2014XVI}:
\def\PR{{\mathcal P_{\mathcal R}}}
\begin{equation} \Delta^2_{\mathcal R}(k)
= A_{\rm s} \left(\frac{k}{k_{\rm p}}\right)^{(n_{\rm s}-1)+\alpha_{\rm s} \log(k/k_{\rm p})/2} \ ,
\label{eq:PRmodel}\end{equation} 
where $A_{\rm s}$ is the primordial amplitude, $n_{\rm s}$ and
$\alpha_{\rm s}$ are the spectral index and the running of the
spectral index, all of which are determined at the pivot scale $k_{\rm p}$. The running of the spectral index can also be equivalently written as:
\begin{equation} 
\alpha_{\rm s} \equiv \frac{{\rm d} n_{\rm s}}{{\rm d} \log{k}} \bigg{|} _{k=k_{\rm p}} .
\end{equation} 
Hence, the matter power spectrum can be written in terms of primordial
quantities as \citep{SmithSimon2025}:
\def\aearly{a_{\rm early}}
\def\aprime{a_{\rm prime}}
\begin{equation}
P_{\rm m}(k,a) = \frac{8\pi^2}{25}\frac{a^2g^2(\aearly,a)}{\Omega_{\rm m}^2}\frac{c^4}{H_0^4}
\,T^{2}(k,a)\,k\Delta^2_{\mathcal R}(k) \ , \label{eq:Pm}
\end{equation} 
where $g(\aearly,a)$ is the growth suppression factor from an early
epoch $\aearly$ to $a$, $T(k,a)$ is the matter transfer function at
epoch $a$ and $c/H_0$ gives the Hubble scale today.
The simulations adopt a fiducial cosmology that is consistent
with the \citet{Planck2014XVI} cosmological parameters -- see
Table\ref{tab:1} for details.

Note that for this parameterisation, taken along with the assumption of
flatness, we can derive a number of other useful cosmological parameters:
the matter density parameter $\Omega_{\rm m}=1-\Omega_{\rm DE}$;
the dimensionless Hubble rate $h =
\sqrt{\left[\omega_{\rm b}+\omega_{\rm c}\right]/\Omega_{\rm m}}$;
the square-root of the variance of matter fluctuations in spheres of radius $R=8 \ h^{-1}{\rm Mpc}$ can also be computed $\sigma_8$ -- 
for details of this see Eq.~(\ref{eq:3.5}), {and varying $A_{\rm s}$ is equivalent to varying $\sigma_{8}^{2}$.}
Through varying each of the 8 cosmological parameters in Eq.~(\ref{eq:2.1})
positively and negatively, additional 16 variational models are
included to investigate the dependence of the nonlinear structure
formation on these parameters. Note that these variations were chosen
to facilitate the construction of derivatives of the observables with
respect to the cosmological parameters and so enable Taylor expansions
of the model around the fiducial point to be constructed. The exact
values of the cosmological parameters used in all models are listed in
Table~\ref{tab:1}. Note that here we correct typos for the parameters
that were presented in Table 1 of \citet{SmithAngulo2018} for models
listed as V5 and V6, which varied the dark energy density (or matter
density).


\begin{figure*}
	\includegraphics[width=2.\columnwidth]{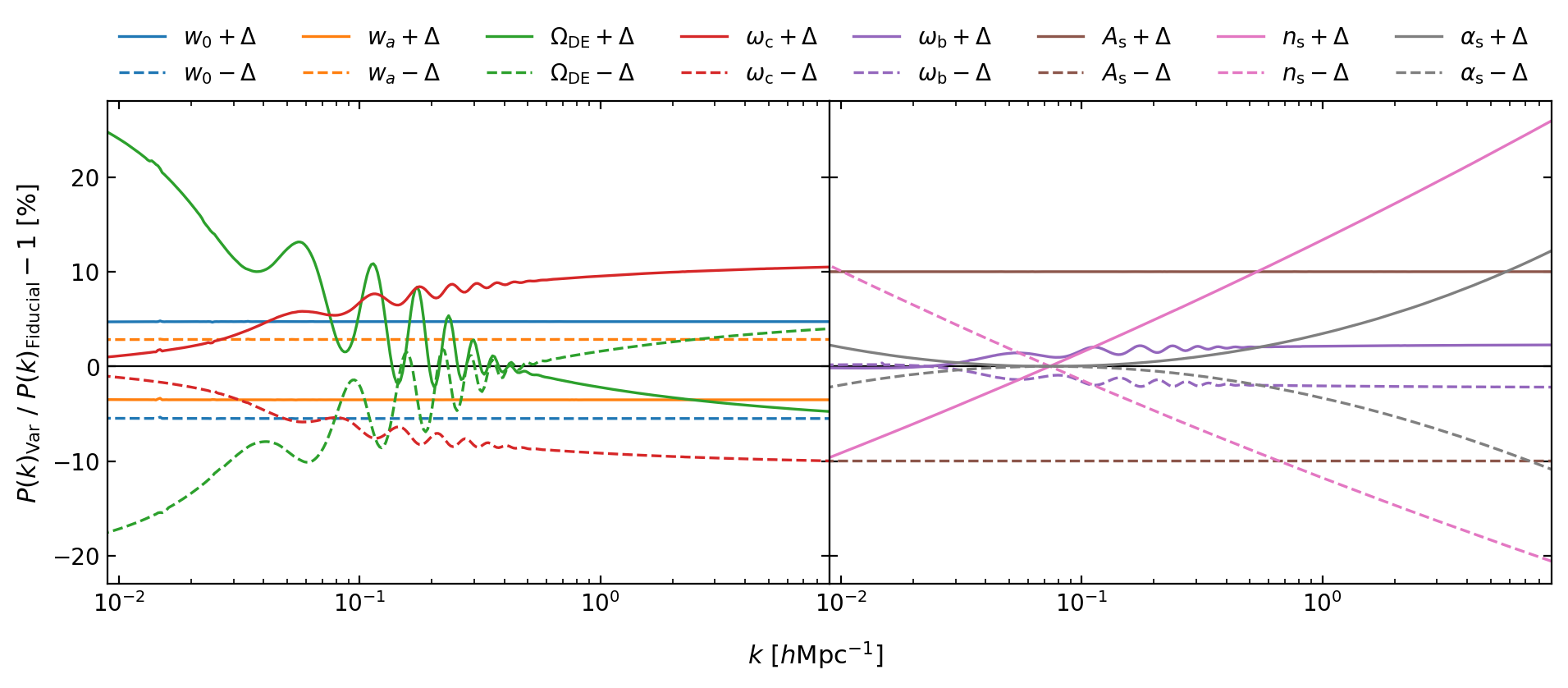}
    \caption{Fractional difference in the linear theory power spectra at $z=0$ between the variational models and the fiducial model. The left panel shows the variations in $w_0, w_a, \Omega_{\rm DE}, \omega_{\rm c}$, and the right panel $ \omega_{\rm b}, A_{\rm s}, n_{\rm s}, \alpha_{\rm s}$. Solid coloured lines represent positive parameter variations, and dashed lines indicate negative variations.}
    \label{fig:Pk_lin}
\end{figure*}


{The linear theory matter power spectra used to generate the initial conditions (ICs) were generated at $z=0$ for all models using the Einstein-Boltzmann solver code {\tt CAMB}. These were then rescaled back to $z=49$ as the input linear power spectra using the linear growth factor described in Section 3.3 in \citet{SmithAngulo2018}, to feed a modified version of the public
{\tt 2LPT} algorithm of \citep{Crocceetal2006}. Figure~{\ref{fig:Pk_lin}} shows the fractional difference in the linear power spectra at $z=0$ between the variational models and fiducial model.} All simulation runs were performed using a special version of the
parallel Tree-PM code {\tt Gadget-3}, which was developed for the
purpose of generating the Millennium-XXL simulation
\citep[][]{Springel2005,Anguloetal2012}. Each run was performed with
$N_{\rm p} = 2048^3$ dark matter particles and $N_{\rm grid} = 2048^3$
PM grid in a box of comoving size $L_{\rm box} = 500 \ h^{-1}
\rm{Mpc}$. The particle mass for all simulations was $m_{\rm p} =
1.240 \times 10^{9} \ h^{-1}M_{\odot}$, with the exception of models
V5 and V6 (see Table~\ref{tab:1}), which have $m_{\rm p} = 1.038
\times 10^{9} \ h^{-1}M_{\odot}$ and $1.442 \times 10^{9}
\ h^{-1}M_{\odot}$, respectively, arising due to the variation of
$\Omega_{\rm m}$. Note that we here correct also the typos of the
particle masses listed in Table 2 of \citet{SmithAngulo2018}.  The
softening length was set to $l_{\rm soft} = 7.5 h^{-1} \rm kpc$, which is
about $3 \%$ of the mean inter-particle spacing $l = L_{\rm box} /
\sqrt[3]{N_{\rm p}}$, which is widely used as the empirical optimal
softening length \citep[][]{Poweretal2003}.

The fiducial model consists of 10 realisations with the same initial
power spectrum, but different realisations of the Gaussian random
field. {We note that the fields are 
true Gaussian random fields, with the initial Fourier modes having random 
phases and the amplitudes being drawn from the Rayleigh distribution. 
For the variational models, only one run was performed, and for each variation 
the initial Gaussian random field generation was set for each variation to match
that of run 1 in the fiducial model with the exception of the power
spectrum model and linear growth rates, which were modified accordingly
for each model.} This was done to help minimise the effect of
large-scale cosmic variance when making comparisons between
variations. Note that we refer to run 1 as the fiducial run hereafter
unless otherwise stated. The haloes were identified using
the on-the-fly FoF algorithm built into {\tt Gadget-3} and was run
with the linking length parameter set to $b = 0.2$. 63 snapshots were
output between $z=49$ and $z=0$ with a hybrid linear-logarithmic time
spacing. In this study we primarily focus on 8 snapshots selected at
$z=0.00$, $0.51$, $0.99$, $1.50$, $2.05$, $3.08$, $4.54$, and $6.75$.


\section{Halo mass functions and universality}
\label{sec:hmf}


\subsection{Background theory}
\label{ssec:theory}

In this subsection we define some key theoretical concepts from
excursion set theory \citep[for a review see][]{Zentner2007}.  To
begin, let us start by defining the number density of haloes with
masses between $m$ and $m+{\rm d}m$ at redshift $z$, $n(m,z){\rm d} m$
where $n(m,z)$ is usually called the differential mass function. The
mass density of all haloes with masses between $m$ and $m+{\rm d}m$ is
then given by $m n(m,z){\rm d}m$. Hence, the fraction of the total
mass in the Universe locked up in these dark matter haloes can then be
defined as:
\begin{equation} f(\nu){\rm d}\nu \equiv \frac{mn(m,z) {\rm
    d}m}{\overline{\rho}_0}\ , \label{eq:3.1} \end{equation} 
where $\overline{\rho}_0$ is the comoving mean matter density of the
Universe. Note that we have written the left-hand-side of this
equation in terms of the variable $\nu$. In this study
we follow the notation in \citet{ShethMoTormen2001} and define this to
be $\nu = \delta^2_{\rm c}(z) / \sigma^2(m)$ \footnote{One should be
careful here on interpreting the literature as some other authors
define $\nu$ as $\delta_{\rm c}(z) / \sigma(m)$, which leads to an
increase of $\nu f(\nu)$ by a factor of 2} (Note, with a small 
misuse of terminology, we shall at times refer to this quantity as the 
`peak-height'). Here $\delta_{\rm c}(z)$
is the time evolving, linear theory, critical overdensity for spherical
collapse of an object at redshift $z$ and $\sigma^2(m)$ is the mass
variance of the density field smoothed by the spherical top-hat window
function at the present day. Note that for CDM power spectra there is
a bijective relation between $\nu$ and $M$ and so one can map back and
forth between them interchangeably.

If we were to invoke the `halo model' framework, then we impose the
condition that all of the mass in the Universe should be locked up in
haloes. This in turn means that if we integrate over all haloes then
the fraction of mass in haloes should give unity:
\begin{equation} \int_{0}^{\infty} f(\nu){\rm d}\nu = \int_{0}^{\infty}\frac{mn(m,z) {\rm
    d}m}{\overline{\rho}_0} = 1\ .
\end{equation} 
We further note that by rearranging Eq.~(\ref{eq:3.1}), to give the fraction
of mass per logarithmic mass interval we can also define the
multiplicity function through\footnote{Note that this is not really
the multiplicity function, as this is actually given by
$m^2n(m)/\rho_0$, which gives the fraction of mass density in haloes
per unit range of $\log m$. However we will persevere with this slight
mislabelling for convenience. }:
\begin{equation} 
    \nu f(\nu) = \frac{m}{\overline{\rho}_0} [mn(m,z)] \frac{{\rm d} \log{m}}{{\rm d} \log{\nu}}\ .
	\label{eq:3.2}
\end{equation} 

The first analytic form for the HMF was developed by
\citet[][]{PressSchechter1974}, with the derivation being put on a
more firm footing by \citet{Bondetal1991}. They found that $n(m, z)$
could be written:
\begin{equation} 
n_{\rm PS}(m, z) = \sqrt{\frac{2}{\pi}} \frac{\overline{\rho}_{0}}{m^{2}}
\frac{\delta_{\rm c}(z)}{\sigma (m)}
  \bigg|\frac{{\rm d} \log{\sigma}}{{\rm d} \log{m}} \bigg|
  \exp{\bigg[ -\frac{\delta_{\rm c}^{2}(z)}{2 \sigma^{2} (m)} \bigg]} \ ,
	\label{eq:3.7}
\end{equation} 
which, when recast in terms of the multiplicity function given by
Eq.~(\ref{eq:3.2}) and making use of $\nu$, leads to the simple functional
form:
\begin{equation} \nu f_{\rm PS}(\nu) = \sqrt{\frac{\nu}{2 \pi}}
\exp{\bigg(-\frac{\nu}{2}\bigg)}\ .\label{eq:3.8}
\end{equation} 
This implies a universal behaviour because the variables that are
dependent on the cosmology and redshift are inserted into $\nu$. Any
change in the cosmology or redshift would only vary $\nu$ rather than
$\nu f(\nu)$.

\citet{LaceyCole1994} and later \citet{ShethTormen1999} subsequently
found that the Press-Schechter model overestimated the abundance of
low-mass haloes and underrepresented the abundance of high-mass haloes
when compared to numerical simulations. {This inspired
\citet{ShethMoTormen2001} to explore
ellipsoidal collapse and more complicated excursion set
scenarios. This opened the path to more complex possibilities for the
analytic form of the HMF. However, when cast in terms of the
multiplicity function, they retained the universal structure,
e.g. \citet{ShethTormen2002} proposed:}
\begin{equation} 
  \nu f(\nu) = A \bigg[1 + \frac{1}{(a \nu)^p} \bigg]
  \sqrt{\frac{a \nu}{2 \pi}} \exp{\bigg(-\frac{a \nu}{2}\bigg)},
	\label{eq:3.9}
\end{equation} 
where $a$ and $p$ were related to the ellipsoidal collapse scenario,
but were taken as free parameters, and $A$ can be determined by
requiring $\int f(\nu) {\rm d}\nu = 1$. This function was shown to
give a significantly better analytic approximation to simulated data.
This approach paved the way for a wide array of numerical studies of
the HMF
\citep{Jenkinsetal2001,White2002,Reedetal2003,Warrenetal2006,Reedetal2007,Lukicetal2007,Tinkeretal2008,Crocceetal2010,Bhattacharyaetal2011,Courtinetal2011,Reedetal2013,Watsonetal2013,Castorianetal2014,Bocquetetal2016,DelPopolo:2017snx,McClintocketal2019,Bocquetetal2020,EuclidXXIV2023} \footnote{Note
that we view studies that have chosen to model the mass function using
a function $f(\sigma^{-1})$ {\em \`a la} \citet{Jenkinsetal2001} as
being approximately equivalent to those modelling $\nu$, since
$\delta_{\rm c}$ is a very weak function of cosmology and since they
encode the same information, but see \citet{Courtinetal2011} for
further discussion of this issue}.
 
As noted earlier, unlike in previous studies, rather than search for a
new functional form for the mass function, our aim is to verify
whether such behaviour exists and if so whether it holds across a
range of cosmological models that includes extensions beyond
$\Lambda$CDM.


\subsection{Ingredients}

In the previous section we defined some important variables and here
we specify them in more detail as this will be needed in what follows.
\begin{itemize}
  \item For the time evolving, linear theory, critical overdensity for
    spherical collapse $\delta_{\rm c}$ we take this to be given by
    the approximation for $\Lambda$CDM as found by
    \citep{KitayamaSuto1996}:
    \begin{equation}\label{eq:3.3}
    \delta_{\rm c} = \frac{3}{20}(12 \pi)^{2/3}[1 + 0.0123 \log_{10} \Omega_{\rm m}(z)] \ .
    \end{equation} 
    In the above expression the matter density parameter evolution is
    given by
    \begin{equation}    
    \Omega_{\rm m}(z) = \frac{\Omega_{\rm m,0}(1 + z)^3}
          {\Omega_{\rm m,0}(1 + z)^3 + (1 - \Omega_{\rm m,0})}\ ,
    \end{equation} 
    where $\Omega_{\rm m,0}$ is the matter density parameter at the
    present time. Note that while we evaluate this expression at the
    necessary redshift, we do not refer to this as the time dependent
    collapse threshold.  Instead we define this following the
    excursion set theory approach and scale this linear theory
    collapse density to the required redshift using the linear theory
    growth factors. This then gives us the barrier for collapse to be
    (note that this is done instead of evolving $\sigma(m)$ with
    time):
    \begin{equation} \delta_{\rm c}(z) = \delta_{\rm c} \frac{D(z=0)}{D(z)}\ , \label{eq:3.4} \end{equation} 
    where $D(z)$ is the linear growth factor (note that it does not
    have to be normalised because we only need its ratio here). We
    have checked that the values of $\delta_{\rm c}$ for the $w$CDM
    cosmology differ weakly (within $1\%$) from the values given by
    Eq.~(\ref{eq:3.3}), regarding the cosmological parameters used in V1,
    V2, V3 and V4 \citep[see][]{Percival2005, Batista2021}. Thus we
    apply the same formula to compute $\delta_{\rm c}$ for all
    models. The computation of $D(z)$ for $w$CDM cosmology is done
    using the publicly available code {\tt CCL}
    \citep[][]{Chisarietal2019}.

  \item We compute the variance of matter fluctuations $\sigma^2(m)$
    smoothed by a spherical top-hat window function, at the present
    day using the expression:
    \begin{equation} \sigma^2(m) = \int \frac{{\rm d}^3k}{(2\pi)^3} P_{\rm
      m}(k) W^2(kR) {\rm d}k,
    \label{eq:3.5}
    \end{equation} 
    where $P_{\rm m}(k)$ is the present day linear matter power
    spectrum given by Eq.~(\ref{eq:Pm}). The top-hat filter in the Fourier
    space is
    \begin{equation}
    W(y) = 3\left[\sin{y} - y \cos y\right]/y^3\ \ ; \ \ y\equiv kR\ \ , 
    \label{eq:3.6}
    \end{equation} 
    where the filter scale $R$ and the mass $m$ enclosed in the sphere
    are related by $m = 4 \pi R^3 \overline{\rho}_0 /
    3$. $\overline{\rho}_{0}$ can be simply obtained via the critical
    density $\rho_{\rm c,0} = 2.7754 \times 10^{11} h^{-1} M_{\odot}
    h^{3 } {\rm Mpc}^{-3}$ and the matter density parameter
    $\Omega_{\rm m,0}$ at present. Note that in computing the numerical 
    integral, we set the lower limit to be
$k_{\rm min}=2\pi / L_{\rm box}$, since in the simulations there is no
power on scales larger than the box wavemode. For the upper
limit we take this to be $k_{\rm max} = \pi/l_{\rm soft}$ where $l_{\rm soft}$ is the softening length. We have checked that the integral has converged to much better than sub-percent precision for these limits regarding the mass range that we consider in this work. We have also explored whether the finite Fourier lattice on which the initial Gaussian random fields were generated could cause a systematic error in the mass function calculations. We found that the idealised matter variance on the lattice agrees with the continuum results to very high precision for mass scales $M\ge M_{\rm Ny}\sim 33 m_{\rm p}$ (see Appendix~\ref{app:Fourierlattice} for details).
\end{itemize}


\subsection{Estimate of the binned mass function}
\label{ssec:estimates}

We now describe how we go about estimating the HMF
${\rm d} n(m)/{\rm d} \log m$ {(equivalently, $mn(m)$, as defined in Eq.~(\ref{eq:3.2}))} and the multiplicity function $\nu f(\nu)$,
as given by Eq.~(\ref{eq:3.2}), from our simulations.

We start by estimating the mass function and we do this by setting up 8
mass bins per decade in logarithmic scale, i.e., $\Delta \log_{10} m =
0.125$. 
We then compute the mass
of each halo by simply multiplying the number of particles in a linked
FoF group by the mass per particle: $M_{\rm FoF} = N m_{\rm p}$. Next we discard the lowest mass haloes from our analysis, due to the
following considerations. \citet{Warrenetal2006} showed that the FoF
halo mass estimates suffer from a systematic bias, which arises due to
the fact that haloes are resolved by discrete particles. They found
that if a halo is sampled by too few particles, then the halo masses
tend to be biased high on average.  \citet{Warrenetal2006} proposed an
empirical formula to correct the particle counts, which has the form
\begin{equation} N_{\rm cor} = N(1-N^{-0.6})\ , \label{eq:3.00} \end{equation} 
where $N_{\rm cor}$ is the estimate of the corrected particle counts
and $N$ is the actual number of particles sampling the halo. This
formula has been widely used in the literature, especially when
treating haloes sampled by several tens of particles \citep[for a
  recent application to the Millennium TNG simulations
  see][]{HernandezAguyaoetal2023}. Subsequent studies by
\citet{Lukicetal2007} and \citet{Courtinetal2011} have also found that
Eq.~(\ref{eq:3.00}) is not a general recipe for mitigating this bias, since
the correction appears to depend on other factors such as halo
concentration, the simulation code used, and the underlying
cosmological model. Hence, its application should be checked case by
case.

Owing to the fact that to correct for this bias robustly would require
a suite of even higher-resolution simulations, we decided to impose
more conservative approach and apply a minimum cutoff in the number
of particles that we would accept haloes to be sampled by. We did not
impose a single cutoff, but tailored these according to our aims. To
be more specific, we adopt a particle cutoff of more than 700
particles for the case when we study the absolute value for the mass
function and its evolution with redshift, which restricts the
systematic bias to below $2\%$ according to Eq.~(\ref{eq:3.00}). By
contrast, when looking at the relative comparison between our
variations in the cosmological model and the fiducial model, we are
more tolerant and impose a cutoff at 150 particles (below $4\%$
according to Eq.~(\ref{eq:3.00}). This is due to the fact that, this
systematic bias should be, roughly speaking, cancelled or at the very
least significantly reduced, when we inspect the ratios of $\nu f(\nu)$
measured from different models. In addition, we
also choose to discard the highest mass bins, removing those with
fewer than 100 haloes per bin. This limits the Poisson errors to
better than $10\%$.


\begin{figure}
  \begin{center}
    \includegraphics[width=1.0\columnwidth]{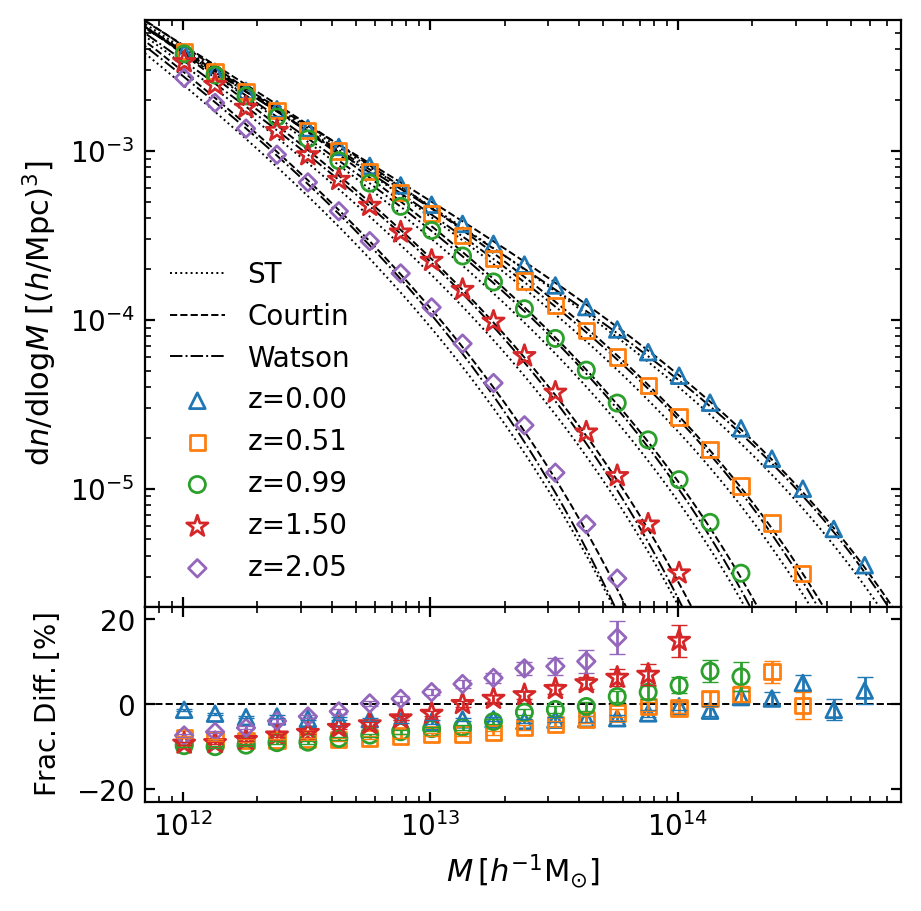}
  \end{center}
    \caption{{\bf Top panel:} Comparison between the measured ${\rm d} n / {\rm d} \log{M}$ as a function of halo
      mass. The data points show the results from the average of the
      10 fiducial cosmology simulations, with the triangles, squares,
      circles, stars, and diamonds denote the results from
      $z=0.00$, $0.51$, $0.99$, $1.50$, and $2.05$. The lines denote the empirical
      models, where the dotted line denotes \citet{ShethMoTormen2001},
      the dashed lines \citet{Courtinetal2011}, and dash-dotted lines
      \citet{Watsonetal2013}. {\bf Bottom panel:} Fractional
      difference of mass functions with respect to the model of
      \citet{Courtinetal2011}. Although only one base line is shown here, the differences are actually with respect to the model at each of the redshifts. Points are as in top panel.}
    \label{fig:0a}
\end{figure}


We next estimate the mass function ${\rm d}n/{\rm d} \log m$ using the
expression
\begin{equation} \frac{ \widehat{{\rm d} n}}{{\rm d} \log{m}} (M_{i}) = \frac{M_{i} \times
  N_{i}}{V \times \Delta M_{i}}\ ,
\label{eq:3.10}
\end{equation} 
where $V$ is the volume of the simulation box, $\Delta M_{i}$ and
$N_{i}$ are the bin width, and number counts of the haloes of the
$i^{\rm th}$ bin, respectively. \citet{Courtinetal2011} found that the
binned mass function is more sensitive to the bin width when using the
averaged mass in each bin to represent $M_{i}$ than using the
bin-centre mass.  Thus here we represent $M_{i}$ with the
bin-centre mass.

Figure~\ref{fig:0a} shows the measured mass functions ${\rm d} n / {\rm d} \log{m}$ as a
function of halo mass and for several redshifts out to $z=2$. For
comparison we also present results for the fitting functions of
\citet{ShethMoTormen2001} (dotted line), \citet{Courtinetal2011}
(dashed line), and \citet{Watsonetal2013} (dash-dotted line). From the
plots it is clear that the abundances evolve strongly with mass and
redshift.

In Appendix~\ref{app:bining} we investigate the effect of bin width on
our binned mass functions, and find that binning leads to an
overestimate of the mass function, especially towards high redshift
and for the higher masses or equivalently rare peaks. As a result, we
also develop a binning correction method, the details of which can be
found there. However, in this work, since we primarily make use of
relative differences in mass functions between different cosmologies
rather than focus on the absolute value, we find that the binning
correction factors also cancel in the ratios.  Finally, we note that,
if one wishes to accurately measure the absolute value of the halo
mass function to high precision, then binning corrections should be
taken into account when less than 10 bins per decade of halo mass are
used.

Next, in order to estimate the multiplicity function $\nu f(\nu)$ we
need to complete the right-hand-side of Eq.~(\ref{eq:3.2}), i.e., determining the quantity ${\rm d} \log{m} / {\rm d} \log\nu$. To do
this, we note that this may also be expressed in terms of $R$ and
$\sigma^{2}$ in the following way:
\begin{equation} 
\frac{ {\rm d} \log m} { {\rm d} \log\nu} = -3 \frac{{\rm d}\log R }{{\rm d} \log \sigma^{2}},
	\label{eq:3.11}
\end{equation} 
which depends only on the present linear matter power spectrum $P_{\rm
  m}(k)$. This can be conveniently estimated from Eq.~(\ref{eq:3.5}) by
either computing the numerical derivative or evaluating the
expression:
\begin{equation} 
\frac{{\rm d}\log \sigma^2(R)}{{\rm d}\log R}
= \int \frac{{\rm d} k^3}{(2\pi)^3}P_{\rm m}(k) W(kR) V(kR) \ . \end{equation} 
where $V(y)$ is defined as:
\begin{equation} 
V(y) \equiv 2\left[9y\cos y+3(y^2-3)\sin y\right]/y^3
\ .\end{equation} 
Note that here we employ the same constraints on the limits of the 
integral as for the case of evaluating $\sigma(m)$.


\begin{figure}
  \begin{center}
    \includegraphics[width=1.0\columnwidth]{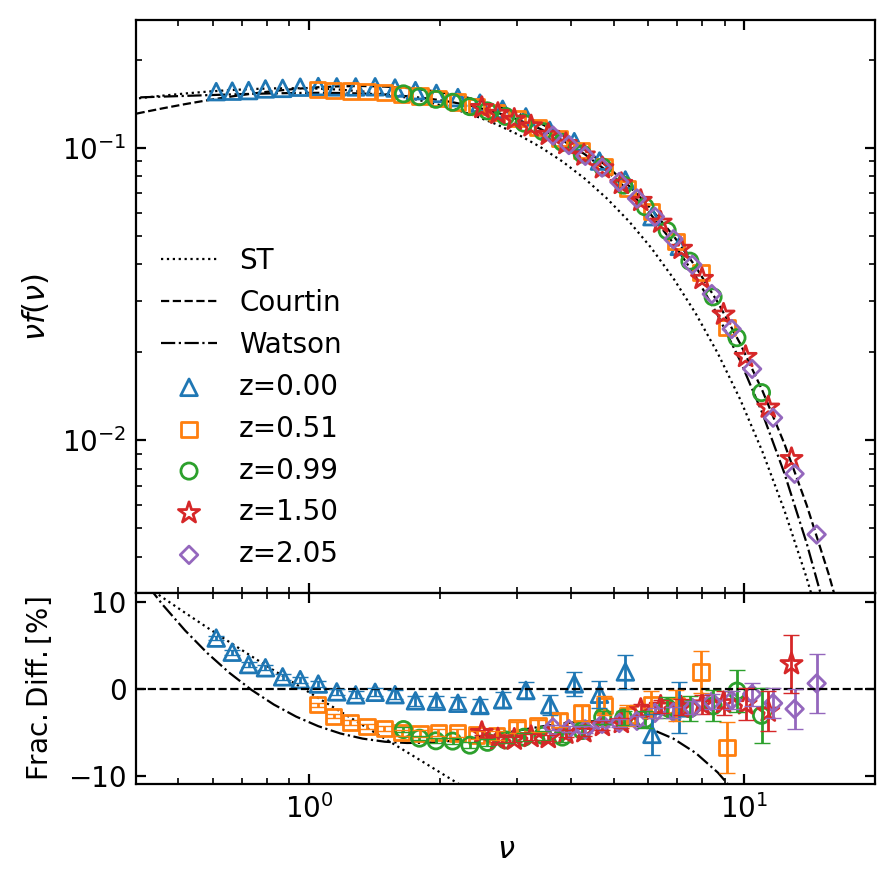}
  \end{center}
  \caption{{\bf Top panel:} Comparison between the measured $\nu
    f(\nu)$ in the fiducial cosmology as a function of $\nu$. The points and lines are as the same as presented in Fig.~\ref{fig:0a}. {\bf Bottom panel:} Ratio of the data
    with respect to the universal form presented in
    \citet{Courtinetal2011}. Points and lines are as before.
  \label{fig:0}}
\end{figure}


Figure~\ref{fig:0} shows a simple comparison between the measured $\nu
f(\nu)$ and the fitting functions in \citet{ShethMoTormen2001} (dotted
line), \citet{Courtinetal2011} (dashed line), and
\citet{Watsonetal2013} (dash-dotted line). The $\nu f(\nu)$ at all 5
redshifts, i.e, $z=0.00$, $0.51$, $0.99$, $1.50$, and $2.05$ are
measured from a total of 10 runs of the fiducial cosmology, amounted
to a volume of $1.25 \ h^{-3} \rm{Gpc}^{3}$. It can be seen that for
the regime $\nu\ge 5$ all of the data points form a similar
locus.  However, for the regime $\nu<5$ the data for $z=0$ are roughly
5\% higher amplitude than the results for higher redshift, {and the data for $z>0$ exhibit an approximate redshift-independent universality.} Thus for
our fiducial simulations we see that universality can only be said to
be approximately true. These findings are in agreement with earlier
results from \citet{Crocceetal2010} and \citet{Courtinetal2011}.

We also note that the \citet{Courtinetal2011} function describes our
$z=0$ data to an accuracy of the order $\sim5\%$, but is
systematically too high for our higher redshift data. On the other
hand, the \citet{Watsonetal2013} function underestimates our low mass
haloes at $z=0$ by $\sim5\%$, but describes the small to medium-$\nu$
regime to high accuracy, though deviates strongly for the high $\nu$
case.


\subsection{Estimate of the error on the HMF}\label{ssec:errors}

We next estimate the errors on the HMF. As was first shown by
\citet{SmithMarian2011} the errors on the binned estimates of the HMF
measured in CDM simulations are not independent, but are in fact
correlated. This correlation arises due to the fact that halo
abundances are modulated by the large scale modes of the power
spectrum. Owing to the fact that we only have 10 realisations of our
fiducial model, and single runs for each of the variational models, to
estimate this covariance, we exploit the estimator described in
Section~5.3 in \citet{SmithMarian2011} \citep[see
  also][]{Crocceetal2010}. The approach is to take advantage of the
fair sample hypothesis and divide the total volume of all the
simulations into a number of smaller cubes, which provides us with a
larger number of quasi-independent volumes but each with a smaller
subvolume: the number of quasi-independent realisations is simply
given by $N_{\rm tot}=N_{\rm runs} \times N_{\rm sub}$ where $N_{\rm
  runs}$ is the number of simulation runs and $N_{\rm sub}$ is the
number of subcubes of each run.  Although these realisations are not
perfectly independent, \citet{SmithMarian2011} checked that there is
no significant impact on the measurement of the HMF covariance.

The covariance of the binned mass function estimates can then be
estimated using:
%
%
\begin{align}
\mathbf{M}x_{ij} &\equiv \left\langle n_i n_j \right\rangle - \left\langle n_i \right\rangle \left\langle n_j \right\rangle \nonumber \\
&= \bigg( \frac{N_{\rm sub}}{V_{\rm tot}} \bigg)^{2} 
\frac{1}{\Delta M_{i} \Delta M_{j}} \frac{1}{N_{\rm tot}} 
\sum_{\alpha,\beta=1}^{N_{\rm tot}} N_{\alpha,i} N_{\beta,j} - n_i n_j,
\label{eq:3.12}
\end{align}

\noindent
where $n_i = N_{i} / (V_{\rm tot} \times \Delta M_{i})$ is the
estimate of the mass function in the $i$th bin averaged over all
subcubes over all realisations, $V_{\rm tot}$ is the total volume of
all runs, $N_{\alpha, i}$ and $N_{\beta, j}$ are the counts of the
$\alpha^{\rm th}$ subcube in the $i^{\rm th}$ bin and the counts of
the $\beta^{\rm th}$ subcube in the $j^{\rm th}$ bin, respectively.
The covariance of the $\nu f(\nu)$ estimates can then be obtained
through repeated application of the factors from Eq.~(\ref{eq:3.2}):
\begin{align}
  \mathbf{N}_{ij} & = \frac{m_i^2m_j^2}{\overline{\rho}^2_0}
  \mathbf{M}_{ij} \left|\frac{{\rm d} \log{m}}{{\rm d}
    \log{\nu}}\right|_{m_i} \left|\frac{{\rm d} \log{m}}{{\rm d}
    \log{\nu}}\right|_{m_j} \ .\label{eq:3.122}
\end{align}
Finally, to examine the strength of the bin-to-bin covariance it is
useful to compute the correlation matrix:
\begin{equation}
    \mathbf{r}_{ij} = \frac{\mathbf{M}_{ij}}{\sqrt{\mathbf{M}_{ii}\mathbf{M}_{jj}}}\ ,
	\label{eq:3.123}
\end{equation} 
whose diagonal elements are unity and whose off diagonal terms are
bounded to $\mathbf{r}_{ij}\in\left[-1,1\right]$. Note that the resultant
correlation matrix is identical if we use either $\mathbf{M}_{ij}$ or
$\mathbf{N}_{ij}$.


\begin{figure}
  \includegraphics[width=1.0\columnwidth]{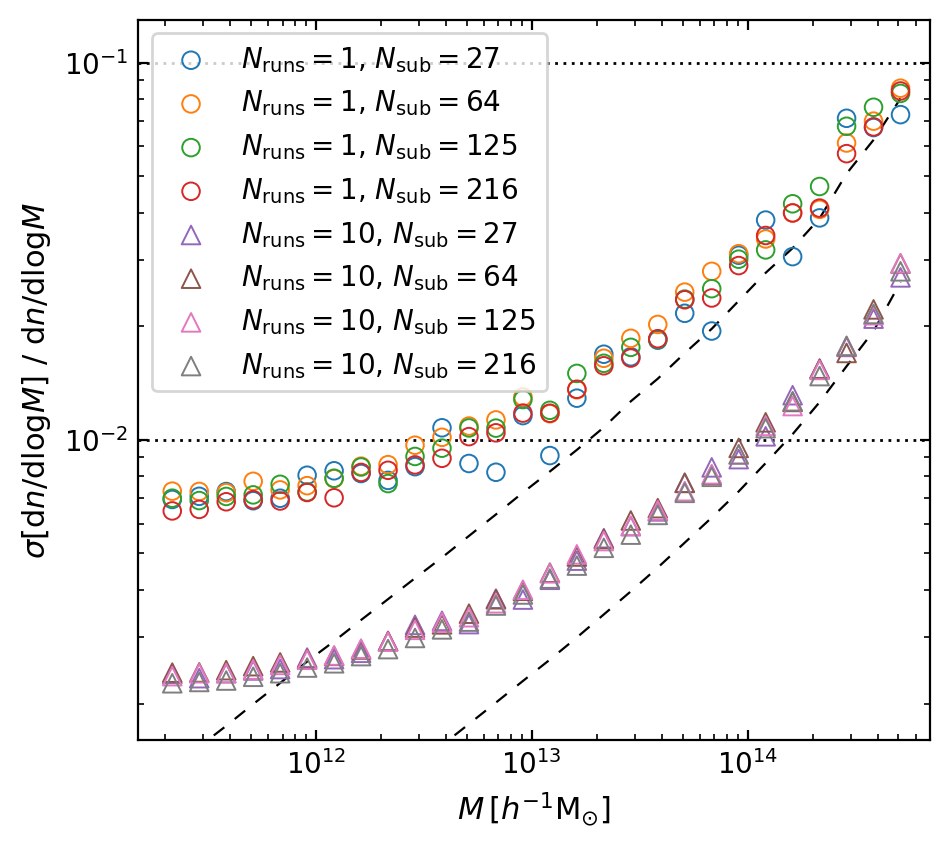}
  \caption{Fractional errors on ${\rm d} n / {\rm d} \log{M}$ at
    $z=0$ for the fiducial model. The results for $N_{\rm runs} = 1$
    (using only the fiducial run) and $N_{\rm runs} = 10$ (using 10
    runs of the fiducial model) are shown here. The blue, orange,
    green and red circles denote the errors estimated using the
    number of divisions of 3, 4, 5 and 6, respectively, on each side
    of simulation box, for $N_{\rm runs} = 1$. The purple, brown,
    pink and grey triangles denote those for $N_{\rm runs} =
    10$. The 2 black dashed lines denote the fractional Poisson
    error for both cases. The 2 dotted horizontal lines mark the
    fractional errors of $10\%$ and $1\%$. The errors weakly depend
    on $N_{\rm sub}$ regarding the same $N_{\rm runs}$.}
  \label{fig:errors}
\end{figure}


Figure~\ref{fig:errors} presents the fractional errors on the halo
mass function estimator, i.e. $\sqrt{\mathbf{M}_{ii}}/n_i$. Here we
show results obtained for the errors for the fiducial model at $z=0$
when we have computed the estimate from Eq.~(\ref{eq:3.12}) with $N_{\rm
  sub} = \{3^3, 4^3, 5^3, 6^3\} = \{27, 64, 125, 216\}$. In the plot
the circled points present the estimates obtained for the first run of
fiducial simulations and triangles give the results for the
combination of all of the 10 fiducial runs. We see that for a single
run the fractional errors are $\lesssim1\%$ for the mass range $M
\lesssim 3\times 10^{12} \ h^{-1} {\rm M}_{\odot}$, but are
$\lesssim10\%$ over the whole mass range considered. On the other hand,
when averaging over all 10 realisations our error estimates are about
3 times smaller and are $\lesssim1\%$ for haloes with $M \lesssim
1\times 10^{14} \ h^{-1} {\rm M}_{\odot}$.

The figure clearly demonstrates that the errors depend only weakly on
our choice for $N_{\rm sub}$ over the whole mass range. In what
follows we choose $N_{\rm sub}=64$ to compute the errors for later
analysis, which corresponds to a size of sample box of $L_{\rm box} =
125 \ h^{-1} \rm{Mpc}$. In addition, the estimated error agrees well
with the fractional Poisson error (dashed lines) at the high-mass end
where the sample size is relatively small. As the number of haloes
grows towards the low-mass bins, the error becomes dominated by sample
variance, which leads to an increase in the statistical error over the
Poisson error \citep{Crocceetal2010,SmithMarian2011}.


\begin{figure}
  \includegraphics[width=1.0\columnwidth]{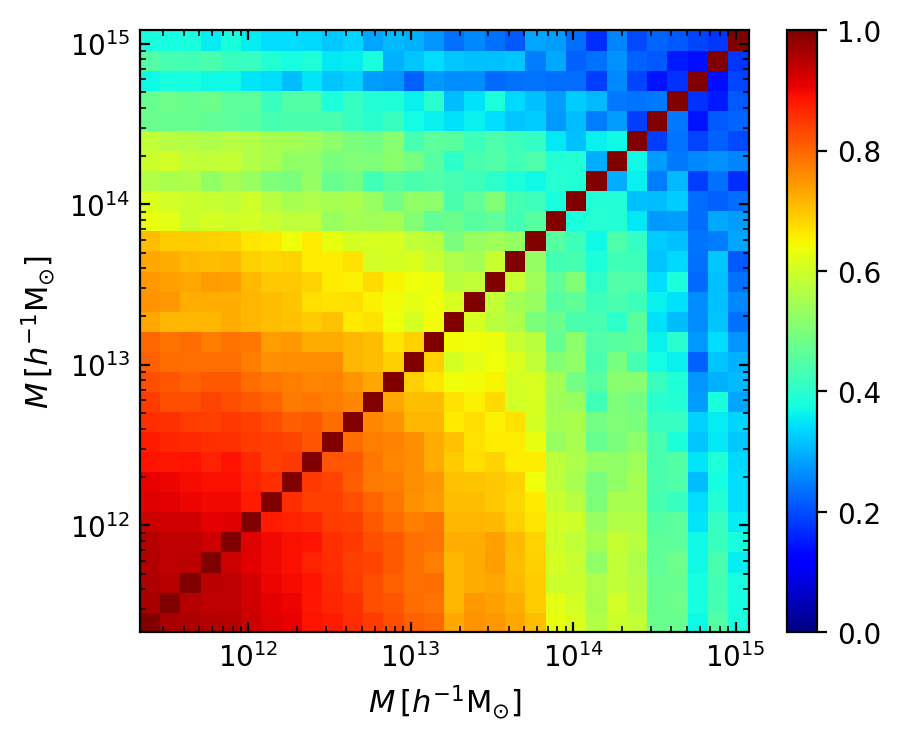}
  \caption{Correlation matrix of the HMF estimated using the sub-cube
    algorithm described in \S\ref{ssec:errors} with $N_{\rm sub}=64$
    for a total of 10 runs in the fiducial cosmology at $z=0$. }
  \label{fig:corr}
\end{figure}


Figure~\ref{fig:corr} shows the correlation matrix $\mathbf{r}_{ij}$
estimated using all 10 runs in the fiducial cosmology at $z=0$, which
amounts to a total volume of $1.25 \ h^{-3} \rm{Gpc}^{3}$. The
estimate is constructed using $N_{\rm sub}=64$ per simulation. The
figure clearly shows that the estimates of low- to intermediate
group-mass bins are highly covariant, i.e., the cross-correlation
coefficient $\mathbf{r}_{ij}\gtrsim 0.5$ for haloes with $M \lesssim
\times 10^{14} \ h^{-1} {\rm M}_{\odot}$). The correlation matrix only
becomes close to diagonal towards the highest mass halo bins with
$M\sim10^{15}\ h^{-1} {\rm M}_{\odot}$. Each of the 10 runs has a
similar correlation matrix but with more noise, the results of which can be
found in the Appendix~\ref{app:corr}.


\begin{figure*}
	\includegraphics[width=2.0\columnwidth]{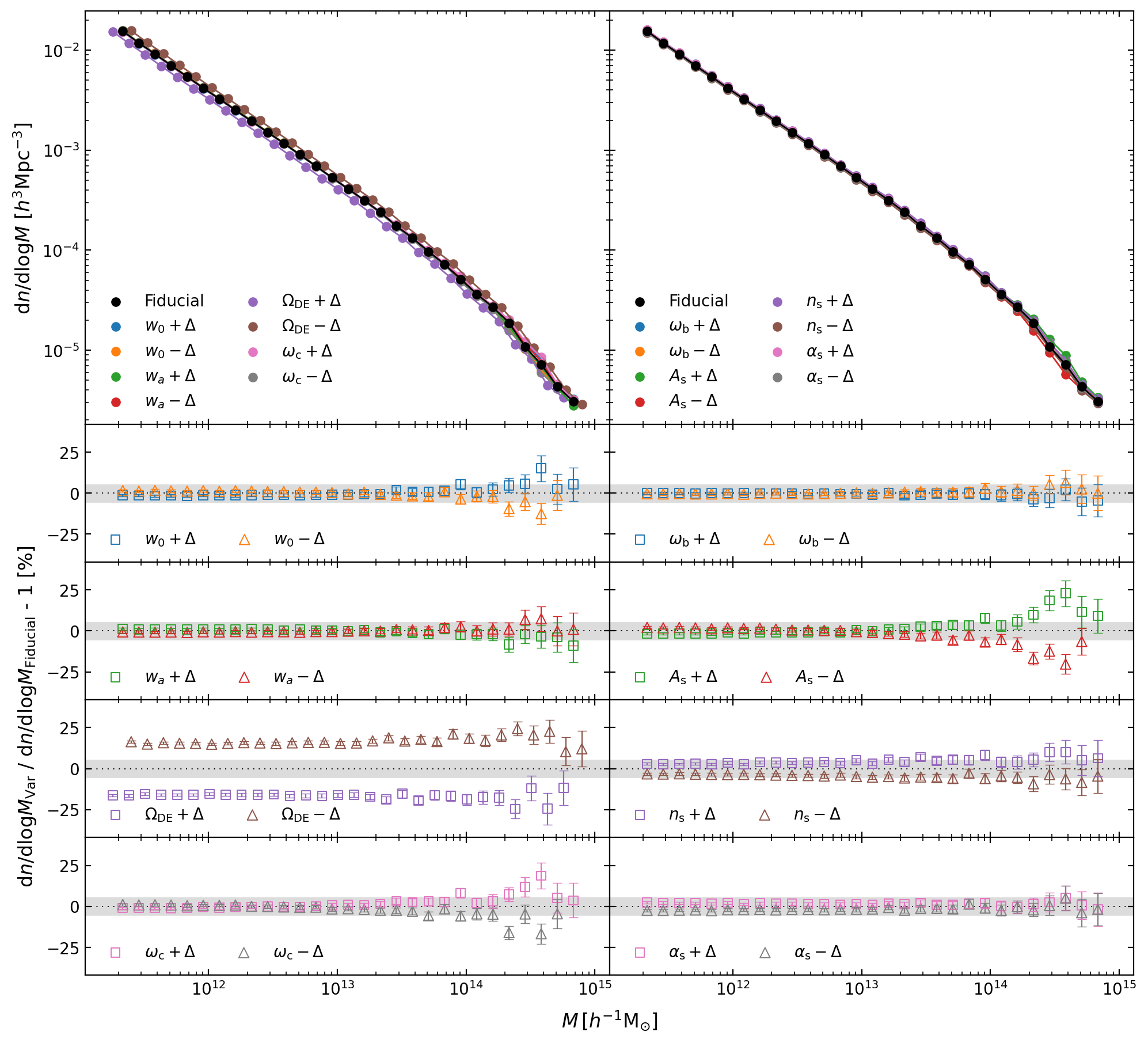}
    \caption{Comparison of ${\rm d} n / {\rm d} \log{M}$ as a function
      of halo masses at $z=0$. The top panels show the ${\rm d} n /
      {\rm d} \log{M}$ measured from the variational runs (coloured solid lines) and fiducial run (black solid
      line), and the other
      panels show the relative difference between them. The coloured
      squares and triangles denote the data for the positive and
      negative models respectively, regarding each cosmological
      parameter. The grey area marks the relative difference of $\pm
      5\%$.}
    \label{fig:2}
\end{figure*}


\begin{figure*}
	\includegraphics[width=2.0\columnwidth]{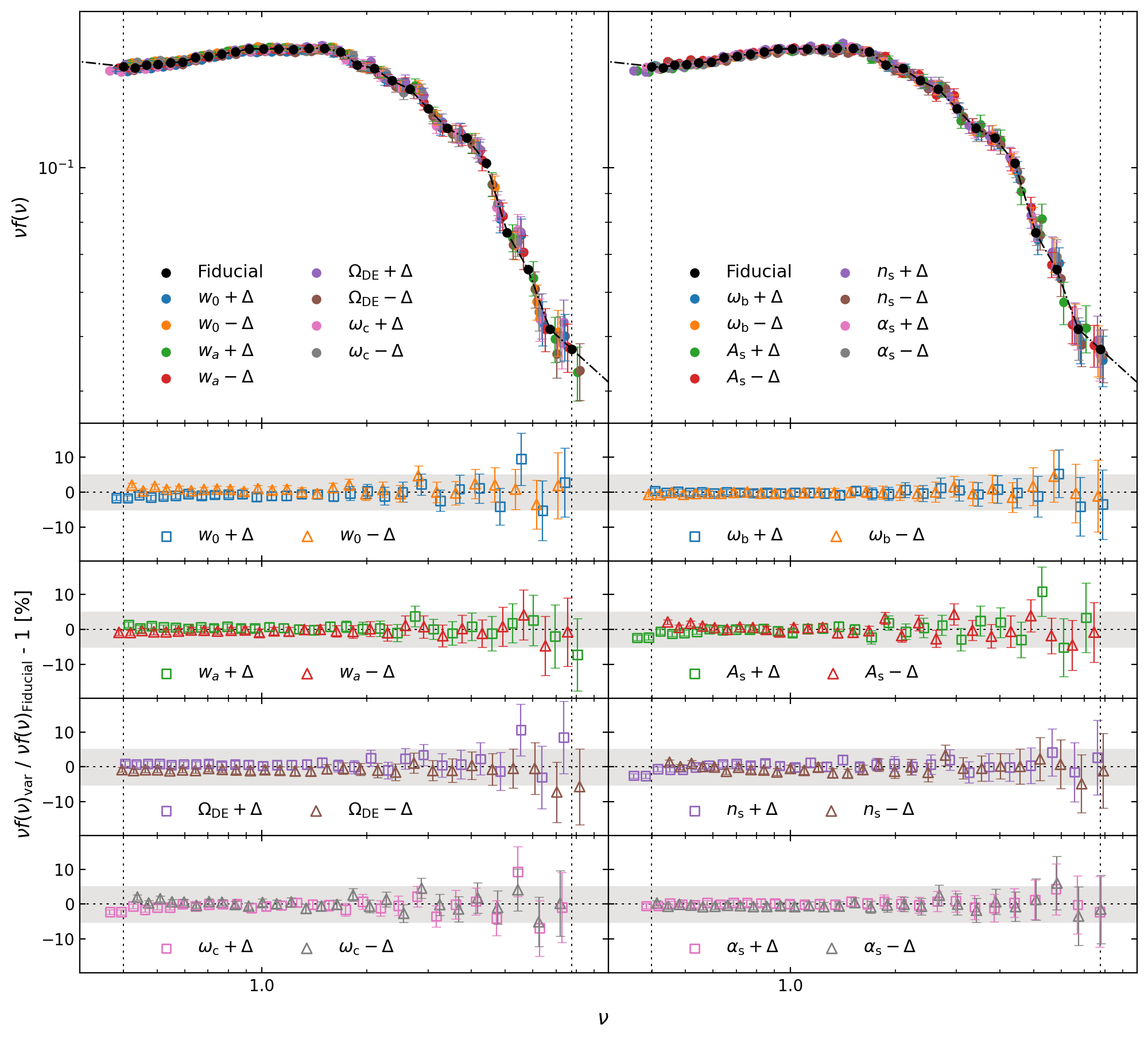}
    \caption{Comparison of $\nu f(\nu)$ as a function of $\nu$ at
      $z=0$. The top panels show the $\nu f(\nu)$ measured from the
      fiducial run (black points) and variational runs (coloured
      points), as well as the assumed universal locus (black
      dash-dotted line) estimated using linear interpolation. The two black
      dotted vertical lines delineate the most left and right black
      points, the locus beyond which is considered to be invalid. The
      rest panels show the relative difference between variational
      runs and the locus. The coloured squares and triangles represent
      the data for respectively the positive and negative variations
      of each of the cosmological parameters. The grey area marks the
      relative difference of $\pm 5\%$. For all variational models,
      most data points reside within the grey area except a few at
      $\nu \gtrsim 5$, which shows an approximate universality of $\nu
      f(\nu)$ at $z=0$ to a different extent exists with respect to
      the considered cosmological parameters.  }
    \label{fig:3}
\end{figure*}


\subsection{{Testing the universality of $\nu f(\nu)$ across cosmological variations}}
\label{ssec:universality}

As was shown in \S\ref{ssec:estimates} universality only approximately
holds for our simulations. However, one of the further aims of this
work is to explore to what extent $\nu f(\nu)$ is approximately
universal and the range of cosmologies for which it can hold,
especially when confronted with extensions beyond $\Lambda$CDM.

To do this then we first measure $\nu f(\nu)$ for our fiducial suite
of simulations at each epoch that we have selected as described in the
previous sections. While averaging over all the simulations would give
us the best estimate of the locus for each redshift, we do not do
this. Instead we take only the results for our run 1 simulation, since
its random phase field matches that of our variation simulations and
so features that can be attributed to large-scale cosmic variance can,
to some extent, be mitigated in the relative difference. At each
epoch, we therefore set up a linear interpolator function of the
measured fiducial locus for $\nu f(\nu)$ {by simply applying the \texttt{scipy.interpolate.interp1d} function \citep[][]{2020Virtanen}} in log-log scale. We then
measure the $\nu f(\nu)$ in the variational runs and compare these
data with the corresponding value obtained from the interpolated
locus. In this way we are able to determine to what extent the
cosmological variations can be said to hold to the locus from the
fiducial run, at that epoch, respectively.

When comparing the data for the variations with the fiducial locus we
apply a reduced chi-square test statistic to quantify the extent of
overall universality for each of the variational runs at different
redshifts. This is given by:
\begin{equation}
        \chi_{n}^{2} = \frac{\chi^2}{n}
    \quad\mathrm{where}\quad
    \chi^2 = \sum_{i=1}^{N} \bigg[\frac{\Delta \nu f(\nu)}{\sigma} \bigg]_{i}^{2}\ ,
	\label{eq:3.13}
\end{equation} 
where $\Delta \nu f(\nu)$ is the difference between the value of $\nu
f(\nu)$ measured in the variational run and the interpolated locus
from the fiducial run. {$\sigma$ is the statistical error that can be obtained through the error on ${\rm d} n / {\rm d} \log{M}$ using Eq.~(\ref{eq:3.2})}, $n$ is the
number of degrees of freedom, i.e., the number of data points minus
the number of fitted parameters. In our case, we do not include any
parameters in the function of the locus. The reduced chi-square is
usually used to describe the goodness-of-fit, with a value of
$\chi_{n}^{2} \lesssim 1$ indicating a good fit since the deviations
of the data are about the same or smaller than the size of their
errors.


\section{Results}
\label{sec:results}

In this section, for reference, we present the results for ${\rm d} n
/ {\rm d} \log{M}$ at $z=0$ for the variational and fiducial runs. We
then present the results of $\nu f(\nu)$ measured from these runs at
the same epoch. Finally, we include the results of $\nu f(\nu)$ at
higher redshifts to show how the universality evolves with redshift
{homing} in on the response with changes to each of the cosmological
parameters varied.


\subsection{Comparison of ${\rm d} n / {\rm d} \log{M}$ at $z=0$}
\label{ssec:hmf-lowz}

Figure~\ref{fig:2}, top panels, shows the absolute mass function ${\rm
  d} n / {\rm d} \log{M}$ estimated for the variational and fiducial
runs at $z=0$. {The left panels show the impact for variations in the equation of
state parameters $w_0$ and $w_a$ and the 
dark energy density parameter $\Omega_{\rm DE}$, along with the variations due to the
physical density of CDM $\omega_{\rm c}$. The right panel shows the variations with the
physical density of baryons $\omega_{\rm b}$, the power spectral amplitude $A_{\rm s}$, its spectral
index $n_{\rm s}$ and its running $\alpha_{\rm s}$.} The bottom panels show
the relative difference between each of the variations and the
fiducial model for a given cosmological parameter. We find that, at
this redshift, the largest differences are for the variation of
$\Omega_{\rm DE}$ {(equivalently a fractional variation of $16 \%$ in $\Omega_{\rm m}$)}, which are greater
than $15 \%$ and span nearly the whole considered mass range ($2 \times 10^{11}
\le M\le 5 \times 10^{14} \ h^{-1} {\rm M}_{\odot}$). For other models,
the deviations are almost within $5 \%$ (grey area) for $M \lesssim 8
\times 10^{13} \ h^{-1} {\rm M}_{\odot}$, but beyond this mass range
growing differences can be observed.

As expected, the variations of some parameters affect ${\rm d} n /
{\rm d} \log{M}$ in a very similar way but with differing strength. For
example, the variations with respect to $w_{0}$, $w_{a}$, $\omega_{\rm
  c}$, $\omega_{\rm b}$, and $A_{\rm s}$ all lead to slightly
positive/negative deviations for small mass scales, and this trend
gradually reverses when considering larger mass scales.
By contrast, the variations with respect to $n_{\rm s}$ and
$\alpha_{\rm s}$ appear to affect the mass functions on all
mass-scales at a similar level. Specifically, we find the
positive/negative variation of them entirely increases/decreases the
halo abundance {over the whole plotted mass range}. Overall, it can be seen that
the mass function, when expressed in the form of ${\rm d}n / {\rm d}
\log{M}$, is quite sensitive with respect to these parameters, and it
is difficult to accurately quantify the effect of the varied
parameters on ${\rm d}n / {\rm d} \log{M}$. This statement is
strengthened when one considers the evolution of the halo abundances
with redshift.


\begin{figure*}
  \includegraphics[width=2.0\columnwidth]{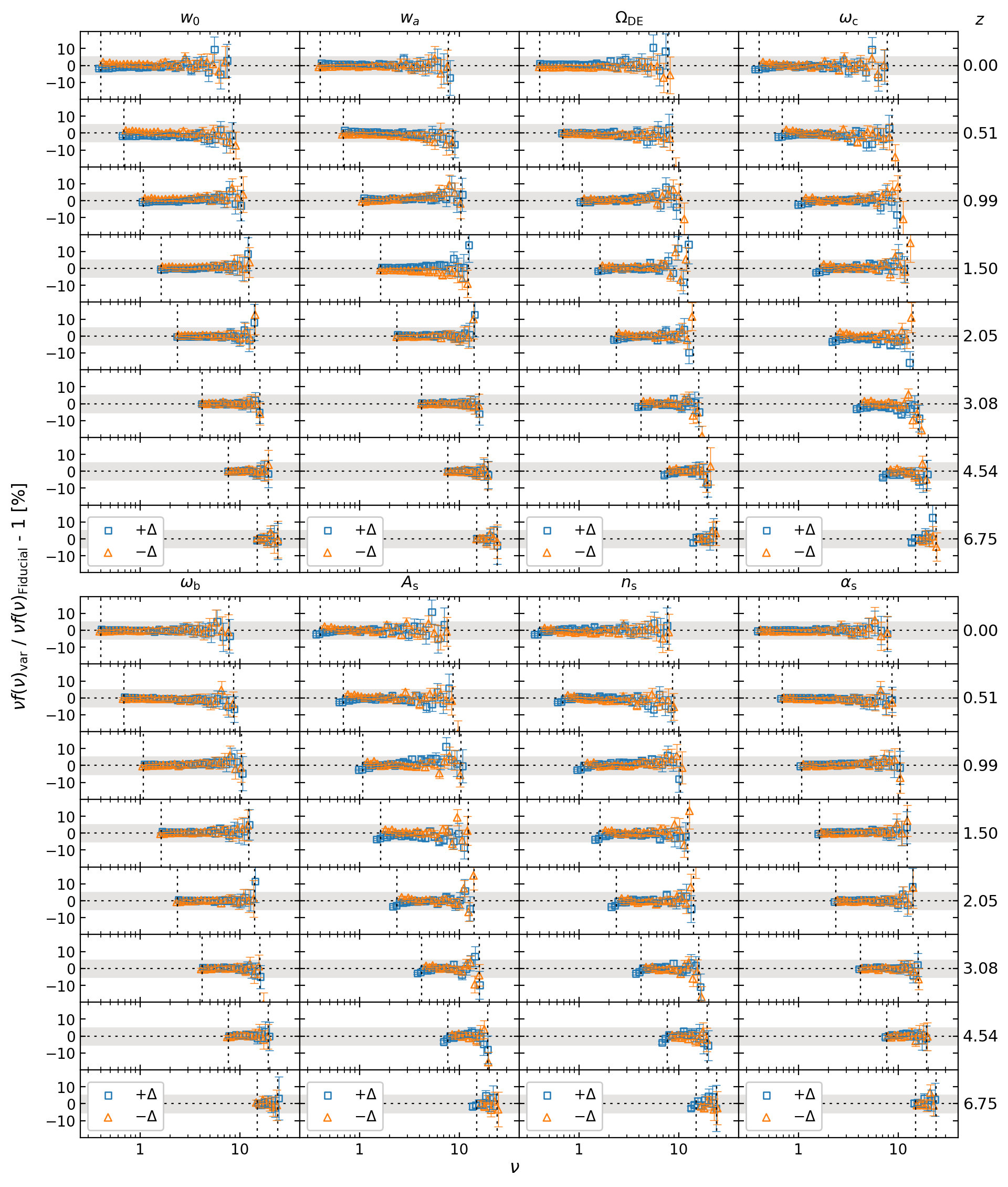}
  \caption{The same as relative difference in Fig.~\ref{fig:3} but
    for more selected redshifts between $z=0$ and $6.75$. Each of
    the columns is for the positive (squares) and negative
    (triangles) variations of the same cosmological parameter, with
    an increase of redshift from the top down. The corresponding
    redshifts are shown on the right side of each row. It can be
    seen that an increase of redshift does not impact the
    approximate universality observed at $z=0$ (shown in
    Fig.~\ref{fig:3}). The $\nu f(\nu)$ plot for these redshifts can
    be found in the Appendix~\ref{app:nufnuhighz} (Fig.~\ref{fig:A1}).  }
  \label{fig:4}
\end{figure*}


\subsection{Comparison of $\nu f(\nu)$ at $z=0$}
\label{Comparison between nufnu at z=0}

Figure~\ref{fig:3} presents the $z=0$ mass functions recast in the
form of $\nu f(\nu)$, with estimates obtained in accordance with the
discussion in \S\ref{ssec:estimates}. The top panels show measurements
from the fiducial (black circles) along with the linear-interpolation model
fit to this data (dash-dotted lines). The results for the variations
in cosmology are presented as the coloured circles. The other panels
show the relative difference between the variational runs and the
linear-interpolation model of the fiducial data. The figure clearly shows that
the relative differences between the variational results and the
fiducial model are almost all within $\pm 5 \%$ over the whole range
of $\nu$ except a few points in the large-$\nu$ tail. However, here
the errors are dominated by the low-counts in the bins, with Poisson
errors being $\Delta \nu f(\nu)/\nu f(\nu)\lesssim 10\%$, given our
cutoff of 100 haloes per bin.

It can also be seen in the $\nu\lesssim0.5$ regime that the ratios for
all variational models are slightly shifted from the universal locus
by $\lesssim2\%$. A possible explanation for this follows from the
discussion in \S\ref{ssec:estimates}: the lower-mass haloes are
sampled by a smaller number of particles and consequently their masses
are overestimated \citep{Warrenetal2006}. In order to investigate this
in more detail, we ran additional 4 simulations with different
resolutions but maintaining all other code parameter settings the
same, i.e., same box size $L_{\rm box} = 500 \, h^{-1} {\rm Mpc}$ but
different particle numbers $N=2048^{3}, 1024^{3}, 512^{3},$ and
$256^{3}$. We found that the bias on the small-$\nu$ tail of $\nu
f(\nu)$ can be significantly removed by using successively
higher-resolution simulations. This implies the possibility that the
universality holds to greater precision for these low-$\nu$ values than
shown in the figure and that it could also possibly be extended to
even smaller-mass haloes. We reserve that investigation for future
work.

We also note that there appears to be features in the shape of the
$\nu f(\nu)$ values estimated from all of the variational simulations
and these features become more accentuated as one considers the
$\nu>1$ data. We have checked that the appearance of these features
can be ascribed entirely to the finite volume effects and the specific
phase distribution of the initial Gaussian random field in Fourier
space. On averaging over all 10 of the fiducial runs the $\nu f(\nu)$
locus that we recover is much smoother. {Additionally, each individual realisation exhibits different features (see Fig.~\ref{fig:D}), which may be one way in which cosmic variance manifests.} However, when we compare the variation
models to the fiducial model, it was important not to use this
averaged $\nu f(\nu)$ spectrum since in this way the sample variance
effects can be better mitigated for. In Appendix~\ref{app:nufnuhighz}
we also show that the features are suppressed when one considers the
$\nu f(\nu)$--$\nu$ locus as a function of redshift (see
Fig.~\ref{fig:A1}).

In Table~\ref{tab:2} we present a more quantitative analysis of the data
and compute the reduced-$\chi_{n}^{2}$ of the variational data as
described in Eq.~(\ref{eq:3.13}). As the table shows, we find that for all
cosmological variations considered at $z=0$ the reduced-$\chi_{n}^{2}$
is $\lesssim 1$ (the largest deviation being for the negative
variation in $A_{\rm s}$).  This indicates that the mass function can
be said to possess a universal form over the mass range $(2 \times
10^{11} h^{-1} {\rm M}_{\odot}<M< 7 \times 10^{14} h^{-1} {\rm M}_{\odot})$. We also found that the values
of $\chi_{n}^{2}$ could be reduced by up to a half of the values
quoted if we tightened the low-mass cutoff mass to include only haloes
with 700 particles or more.


\begin{table*}
  \centering
  \caption{Reduced Chi-square $\chi_{n}^2$ for each of the
    variational runs at different redshifts. Except the first
    column that denotes redshift, every two columns denote the
    positive and negative variations of each cosmological
    parameter. A value of $\chi_{n}^2 \lesssim 1$ indicates a
    good universality of $\nu f(\nu)$.}
  \label{tab:2}
  \begin{tabular}{c|c|c|c|c|c|c|c|c}
    \hline
    \hline
    $z$ & $w_{0}+\Delta$ & $w_{0}-\Delta$ & $w_{a}+\Delta$ & $w_{a}-\Delta$ & $\Omega_{\rm DE}+\Delta$ &
    $\Omega_{\rm DE}-\Delta$ & $\omega_{\rm c}+\Delta$ & $\omega_{\rm c}-\Delta$ \\
    \hline
    $0.00$ & $0.966$ & $0.949$ & $0.483$ & $0.240$ & $0.576$ & $0.785$ & $0.592$ & $0.862$ \\
    
    $0.51$ & $1.949$ & $0.783$ & $0.633$ & $0.611$ & $0.313$ & $0.403$ & $0.933$ & $1.166$ \\
    
    $0.99$ & $0.260$ & $1.448$ & $1.149$ & $0.924$ & $0.567$ & $0.856$ & $0.513$ & $1.137$ \\
    
    $1.50$ & $0.142$ & $0.745$ & $0.436$ & $1.235$ & $0.737$ & $1.420$ & $0.614$ & $1.627$ \\
    
    $2.05$ & $0.170$ & $0.221$ & $0.232$ & $0.133$ & $0.852$ & $0.947$ & $2.246$ & $1.059$ \\
    
    $3.08$ & $0.105$ & $0.114$ & $0.115$ & $0.048$ & $0.708$ & $0.874$ & $1.997$ & $0.993$ \\
    
    $4.54$ & $0.057$ & $0.066$ & $0.067$ & $0.125$ & $0.269$ & $0.439$ & $0.753$ & $0.548$ \\
    
    $6.75$ & $0.118$ & $0.291$ & $0.062$ & $0.178$ & $0.233$ & $0.220$ & $0.641$ & $0.246$ \\	
    \hline
    \hline
    $z$ & $\omega_{\rm b}+\Delta$ & $\omega_{\rm b}-\Delta$ & $A_{\rm s}+\Delta$ & $A_{\rm s}-\Delta$ &
    $n_{\rm s}+\Delta$ & $n_{\rm s}-\Delta$ & $\alpha_{\rm s}+\Delta$ & $\alpha_{\rm s}-\Delta$ \\
    \hline
    $0.00$ & $0.107$ & $0.139$ & $0.517$ & $1.122$ & $0.422$ & $0.676$ & $0.097$ & $0.283$ \\
    
    $0.51$ & $0.190$ & $0.328$ & $0.740$ & $1.278$ & $0.596$ & $0.899$ & $0.195$ & $0.407$ \\
    
    $0.99$ & $0.417$ & $0.308$ & $1.304$ & $1.569$ & $1.080$ & $0.824$ & $0.435$ & $0.333$ \\
    
    $1.50$ & $0.393$ & $0.191$ & $2.205$ & $1.605$ & $0.800$ & $0.759$ & $0.341$ & $0.187$ \\
    
    $2.05$ & $0.110$ & $0.139$ & $0.491$ & $1.313$ & $0.441$ & $0.563$ & $0.213$ & $0.148$ \\
    
    $3.08$ & $0.118$ & $0.153$ & $0.620$ & $1.007$ & $0.431$ & $0.546$ & $0.199$ & $0.123$ \\
    
    $4.54$ & $0.129$ & $0.129$ & $0.293$ & $0.415$ & $0.525$ & $0.309$ & $0.220$ & $0.264$ \\
    
    $6.75$ & $0.101$ & $0.215$ & $0.470$ & $0.375$ & $0.498$ & $0.498$ & $0.265$ & $0.422$ \\	
    \hline
    \hline
    
  \end{tabular}
\end{table*}


\subsection{Comparison of $\nu f(\nu)$ at higher redshifts}
\label{ssec:nufnuevolve}

In Figure~\ref{fig:4} we explore the redshift evolution of the
universal multiplicity function $\nu f(\nu)$ ratio of the variational
runs with respect to the fiducial model.  The top panel shows results
for the parameters $w_0$, $w_a$, $\Omega_{\rm DE}$ and $\omega_{\rm
  c}$, whereas the bottom panel shows results for $\omega_{\rm b}$,
$n_{\rm s}$, $A_{\rm s}$ and $\alpha_{\rm s}$. Going from top to
bottom the subpanels show the $\nu f(\nu)$ ratio results for the
redshifts $z=\{0.0,0.51,0.99, 1.5, 2.05, 3.08, 4.54, 6.75\}$.  We
notice that the approximate universality observed at $z=0$ holds to
better than $\lesssim5\%$ for higher redshifts. The possible outliers
being for $A_{\rm s}$ and $\omega_{\rm c}$. These qualitative
observations are more quantitatively demonstrated in the results from
Table~\ref{tab:2}. This shows that most models share a similar
tendency that as redshift goes up, the values of $\chi_{n}^{2}$
firstly increase and then start to drop at $z > 1$ or $1.5$, which
possibly indicates that universality holds better at low and high
redshifts.

In more detail, for the parameters $\omega_{\rm b}$ and $\alpha_{\rm
  s}$ we find that the $\nu f(\nu)$ locus at higher redshifts is
incredibly well described by the locus from the fiducial
model. Looking at the simulations that vary the dark energy-related
parameters $w_{0}$ and $w_{a}$, we again see good agreement for all
relevant redshifts but with a potential deviation for $0.5 \lesssim z
\lesssim 1.5$, where the universality appears to be slightly broken.
At higher redshifts this deviation is no longer visible. For
$\Omega_{\rm DE}$, or equivalently $\Omega_{\rm m}$, the points become
very close to the locus in the small- and medium-$\nu$ regimes at $z
\gtrsim 0.5$, whereas moderate deviations can be observed with the
large-$\nu$ tail at each of the redshifts. Likewise, regarding
$\omega_{\rm c}$ and $A_{\rm s}$, a relatively good match can be seen
with the small and medium scales of $\nu$ at all redshifts, but the
points in the large-$\nu$ regime severely deviate from the
locus. Interestingly, a similar trend of deviations can still be
observed at some of the higher redshifts for both parameters. It is
very likely that these variations have an equivalent influence on $\nu
f(\nu)$. looking at $n_{\rm s}$, we see a good match to the fiducial
locus, but again with some possible deviations around $z=1$.


\section{Discussion}\label{sec:discussion}

\subsection{Utility of a universal FoF halo mass function}

In a number of studies focused on cluster cosmology the utility of the
FoF HMFs has been foregone in favour of studying haloes
identified through either the spherical overdensity algorithm
\citep{Tinkeretal2008,McClintocketal2019,Costanzietal2019a,Costanzietal2019b,Bocquetetal2020}. The objections mostly focus on the notion that the haloes selected by a spherical overdensity algorithm are more physically meaningful
quantities since one can link the objects found to the virialisation
overdensities resulting from the spherical collapse model, whereas the
FoF algorithm with fixed $b=0.2$, links to haloes of varying
overdensity \citep{Courtinetal2011,Moreetal2011}. Further, it is
speculated that the cluster observable--halo mass distribution
function can be more readily calibrated. These arguments can be
countered by noting that it is well known that the condensations found
in simulations are not spherical \citep[for example
  see][]{Courtinetal2011,Moreetal2011}. In fact gravitational collapse
leads to a distribution of prolate and oblate ellipsoids
\citep{Bardeenetal1986,ShethMoTormen2001, JingSuto2002}. In this
sense, one should consider spherical haloes as being {\em less}
physically meaningful. {Treating ellipsoidal haloes as spherical will
lead to underestimates of the mass \citep{Despali2013}}. Secondly, owing to the fact that
the cluster observable--halo mass relation requires some detailed
knowledge of galaxy formation physics and observational selection
effects, there is no one-to-one correspondence between any 3D halo
identified in simulations and any cluster observable. These have to be
fully calibrated. With the FoF algorithm, one can at least be sure
that one is not missing mass, which can happen in SO identified
haloes.

Ultimately, for accurate cluster cosmology, one simply needs to
carefully model either of the conditional probability density
functions $P(X|M_{\rm SO})$ or $P(X|M_{\rm FoF})$, where $X$ is a
given cluster observable.  However, when it comes to making accurate
predictions for the cluster counts, these will be obtained by
convolving the above PDFs with the respective model for the mass
functions. Our findings in this work suggest that the modelling of the
$mn(m)$ for FoF haloes is simplified by the ability to exploit the
near universality shown, especially when one considers extensions
beyond the simple $\Lambda$CDM model. {All of this leads us to the
recommendation that it is $P(X|M_{\rm FoF})$ one could consider expending
resources to calibrate. Additionally, we do not think of FoF haloes as the only path to a universal mass function, what we want to do is to provide motivation for finding a good candidate that can be used as a base model around which model variations are minimal, and so making the emulation more accurate and giving potential wider parameter coverage. We believe that FoF and SO haloes have both their advantages and disadvantages, it would be better that people can have more options of halo definitions to choose according to their research aims.}


\subsection{Parameter space coverage}

\begin{figure}
  \includegraphics[width=1.0\columnwidth]{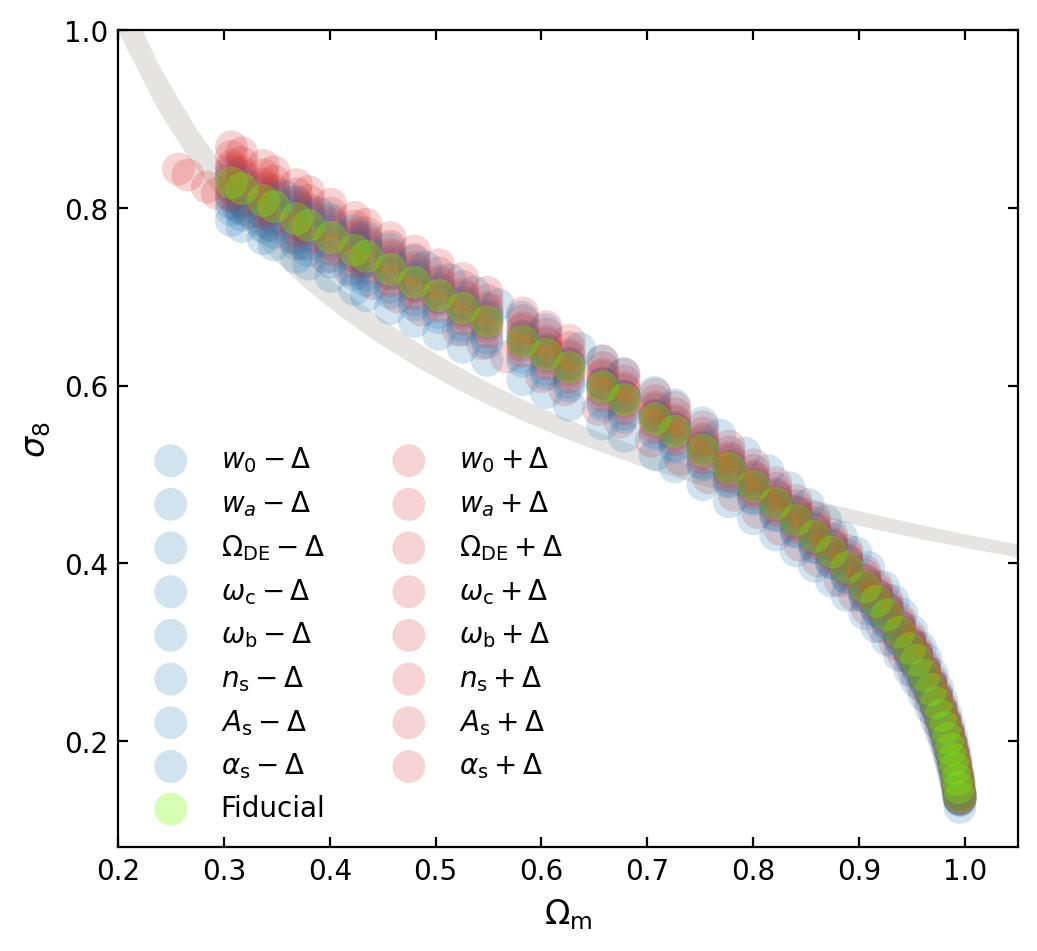}
  \caption{The values of $\sigma_{8}-\Omega_{\rm m}$ covered by all
    snapshots between $z=0$ and $6.75$ of all cosmological models
    used in this work. Red, blue, and green transparent circles respectively denote the positively, negatively varied models, and the fiducial model. The points of each model starts from upper
    left with $z=0$ to lower right with $z=6.75$. The grey area
    shows the best-fit $\sigma_{8}-\Omega_{\rm m}$ degeneracy
    obtained from cluster abundance in \citet{Abdullahetal2020}. A
    good overlap can be seen with $\Omega_{\rm m} \lesssim 0.4$,
    which indicates where our results could apply to in the
    $\sigma_{8}-\Omega_{\rm m}$ plane.  }
  \label{fig:5}
\end{figure}

It is well known that when constraining the cosmological parameters
from the HMF there is a degeneracy between
$\sigma_{8}$--$\Omega_{\rm m}$. It is therefore interesting to examine
where on this plane our results of near universality hold. To do this,
we can think of the results from higher redshift snapshots as defining
a different world model with a higher $\Omega_{\rm m}$ content and a
corresponding lower $\sigma_{8}$. Thus we can use these results to
cover a wider range of locations on the $\sigma_{8}$--$\Omega_{\rm m}$
plane than one would expect from simply examining the $z=0$ models
alone.

Figure~\ref{fig:5} shows the collection of points in the
$\sigma_{8}$--$\Omega_{\rm m}$ plane that are covered by 47 of the
snapshots that we analysed between $z=0$ and $6.75$ in all of the
cosmological models used in this work. In the plot we also compare our
parameter space coverage with the best-fit $\sigma_{8}-\Omega_{\rm m}$
degeneracy obtained from cluster abundance in
\citet{Abdullahetal2020}. We see that there is a good overlap between
our sampled points and this curve for $\Omega_{\rm m} \lesssim 0.4$.
However, we expect that our results should hold well beyond the locus
of points shown. It is our intention to explore this further in a
future study.


\subsection{Caveats to the analysis}

Some caveats to over-interpreting our findings are that our approach
of examining the ratios has neglected to account for the noise in the
fiducial model and the errors that are shown are those for the
variational models alone. Furthermore, in our analysis of the
reduced-$\chi^2$ as described in Eq.~(\ref{eq:3.13}), we are also neglecting
the errors in the interpolated locus of the fiducial model. A better
approach would be to take account of these errors when performing the
test. This might possibly be done by building a Gaussian process model
for the fiducial model and then use that to forward model predictions
and errors for the actual bins used in the variational models
\citep[see][for discussion of the emulation]{McClintocketal2019}.  A
further source of improvement would be to consider the inclusion of
the off diagonal components of the covariance matrix. However, this
would likely require a larger suite of simulations in order to minimise
the errors induced in computing the inverse of the matrix. We leave
these investigations for future work.


\section{Conclusions}\label{sec:conclusions}


In this paper we have investigated the universality of the abundance
of dark matter haloes in the $\Lambda$CDM and $w$CDM frameworks and
have also explored the impact of modifications of the primordial
physics through variations with respect to the amplitude, spectral
index of the primordial power spectrum, and along with a possible
running of the primordial power spectral index. For our exploration,
we made use of the D\"ammerung simulations, a suite of high-resolution
($N=2048^3$) dark matter only cosmological $N$-body simulations,
carried out by \citet{SmithAngulo2018}. The simulations consisted of
10 runs of a fiducial model that adopted the Planck 2014 cosmology and
16 variational runs that positively and negatively varied each of the
8 cosmological parameters around the fiducial model. We selected
haloes using the FoF($b=0.2$) algorithm. This choice owed to the fact
that previous studies had demonstrated a near universality for these
objects and due to the advantage of computational efficiency and
scalability.

In order to test the universality $\nu f(\nu)$ we built a linear
interpolation model of the results from our fiducial run. On
comparison of the $\nu f(\nu)$ measured from the variational runs with
this model, we were able to assess the validity of the putative
universal behaviour with respect to the 16 variations. We also
selected a number of snapshots with a coverage of redshift
$z\in[0,6.75]$ to inspect evolution with redshift. To make quantitative
statements, we performed a reduced-$\chi^2$ test, for which we applied
the subcubing approach introduced in \citet{SmithMarian2011} to
compute the covariance matrices of the HMFs and so estimate
statistical errors. We summarise our findings as follows:

\begin{itemize}

\item Considering the resolution and volume of each of the D\"ammerung
  simulations, we found that reliable results were obtained for the
  haloes in the mass range $(1.0 \times 10^{12} h^{-1} {\rm M}_{\odot} \leq M \leq 7.0
  \times 10^{14} h^{-1} {\rm M}_{\odot})$. Comparing to measurements of
  ${\rm d} n /{\rm d} \log{M}$ from our fiducial model, we found that at $z=0$ the
  variations of the cosmological parameters $w_{0}$, $w_{a}$,
  $\omega_{\rm c}$, $\omega_{\rm b}$ and $A_{\rm s}$ mainly impacted
  ${\rm d} n / {\rm d} \log{M}$ for $M \gtrsim 10^{14} \ h^{-1}
  M_{\odot}$, whereas for mass scales $M \lesssim 10^{14} \ h^{-1}
  M_{\odot}$ the effects were more modest. On the other hand, the
  variations of $\Omega_{\rm DE}$ ($\Omega_{\rm m}$) led to strong
  changes in ${\rm d} n / {\rm d} \log{M}$ over the whole mass
  range. By contrast, variations arising due to modifications of
  $n_{\rm s}$ and $\alpha_{\rm s}$, led to moderate effects on all
  scales.

\item For the redshift range $(0 \lesssim z \lesssim 6.75)$, we found
  that the relative differences between the $\nu f(\nu)$ across all
  variations at each redshift were better than $\pm 5 \%$ over the
  whole range of $\nu$ considered, except for a few points in the
  large-$\nu$ tail. However, these excursions were likely due to
  sample variance.  We applied the reduced-$\chi_{n}^{2}$ test to
  quantify the extent of the overall universality and found for the
  $z=0$ data values of $\chi_{n}^{2} \lesssim 1$, indicating good
  approximate universality at that redshift across the models we
  considered. For higher redshifts, the reduced-$\chi_{n}^{2}$
  slightly increased and then dropped for $z \gtrsim 1.5$, which
  suggested that approximate universality holds more strongly at both
  low and high redshifts.  However, some of the deviations we noted
  could be due to the fact that we did not fold in the sample and
  Poisson errors of the fiducial model into the ratio, which softens
  our conclusions at these redshifts and leaves room for further
  improved agreement between models.

\item Amongst all of the parameter variations that we considered, the
  universality of $\nu f(\nu)$ was found to hold to very high
  precision for $\omega_{\rm b}$ and $\alpha_{\rm s}$ across all $\nu$
  and $z$ considered. The variations of all of the other parameters
  $w_0$, $w_a$, $\Omega_{\rm DE}$, $\omega_{\rm c}$, $n_{\rm s}$ and
  $A_{\rm s}$ was similarly good at low and high redshift, but with
  hints of a breakdown occurring at intermediate redshifts. However,
  all departures were $\lesssim5\%$.

\item We noticed that for nearly all of the cosmological models
  considered the points in the small-$\nu$ regime (corresponding to
  haloes with mass scales $M \lesssim 10^{12} h^{-1} {\rm M}_{\odot}$) were slightly
  shifted from the universal locus. We concluded that this was likely
  caused by the systematic bias in the mass estimates for FoF haloes
  sampled by finite numbers of particles. This implied that the good
  approximate universality found here could actually be extended to an
  even wider range of halo masses. However, to confirm this, would
  require a suite of $N$-body simulations of even higher resolution
  and so we leave confirmation of this to future work.
  
\end{itemize}

In recent years there has been a rapid increase in the methods of
identifying and classifying dark matter haloes in numerical
simulations and in observational data. A number of theoretical studies
have followed the path of studying haloes identified through the SO
algorithm.  However, it has been widely confirmed that the HMFs for
these objects are far from a universality except using the virial overdensity and thus require a large amount
of simulations to accurately calibrate their abundance as a function
of cosmological model and over time. By contrast, haloes identified
through the FoF algorithm exhibit approximate universality with
respect to a wide set of cosmological parameters. The advantage of
this is that only several mass functions need to be accurately
parameterised so that they can be deployed for likelihood modelling in
cluster abundance studies.

In this work, we have completed a first step to producing an improved
model for the HMF. We have quantitatively shown that
FoF haloes do have the potential for building such models with the
minimum required sets of numerical simulations.  Our next step will be
searching for an effective form of the universal $\nu f(\nu)$ that can
provide percent-level constraints. The ultimate goal, of course, is to
produce a theory forward modelling pipeline for use in the
likelihood function for observing cluster counts binned in terms of
observable mass proxies for upcoming galaxy surveys such as Euclid. To
achieve this, one needs to understand the conditional probability
density of obtaining the observable given the underlying halo.

In this study we have also neglected how modifications of the baryonic
physics may impact the universality of the mass function. Nor have we
included the study of massive neutrinos or modifications of the dark
matter particle modelling. These we shall reserve for future study.


\section*{Acknowledgements}

We would like to thank Adrian Jenkins and Ravi Sheth for their careful reading of the manuscript and useful suggestions, which have improved the paper. RES acknowledges support from the Science and Technology Facilities Council (grant ST/X001040/1). YL thanks Baojiu Li for useful discussion. This work used the DiRAC@Durham facility managed by the Institute for Computational Cosmology on behalf of the STFC DiRAC HPC Facility (www.dirac.ac.uk). The equipment was funded by BEIS capital funding via STFC capital grants ST/K00042X/1, ST/P002293/1, ST/R002371/1 and ST/S002502/1, Durham University and STFC operations grant ST/R000832/1. DiRAC is part of the National e-Infrastructure.  We acknowledge that the results of this research have been achieved using the PRACE Tier-0 Research Infrastructure resource SuperMuc based in Garching Germany at the Leibniz Supercomputing Centre (LRZ) under project number 2012071360.


\section*{Data Availability}

Simulation data used in this work can be made available upon request to the authors.



\bibliographystyle{mnras}
\bibliography{massfunction} 



\appendix


\section{Effect of the Fourier lattice on the mass variance}
\label{app:Fourierlattice}


\begin{figure}
    \includegraphics[width=1.\columnwidth]{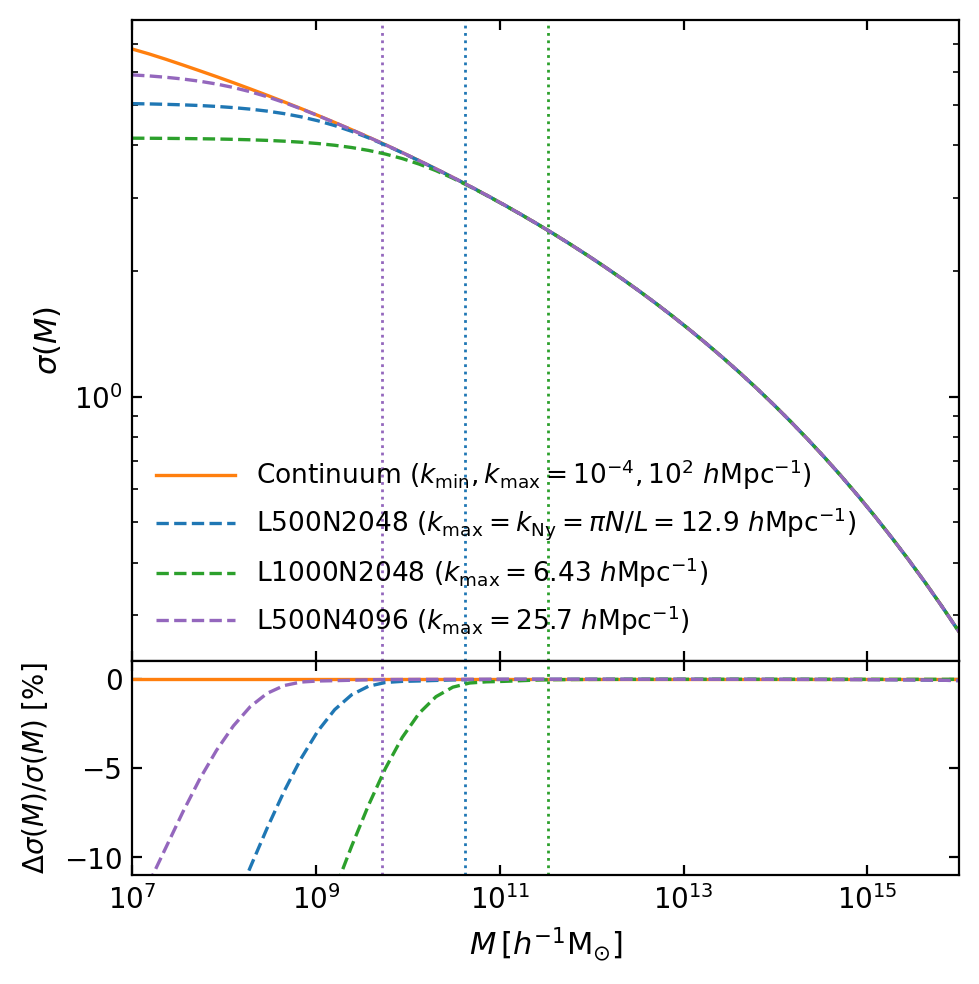}
    \caption{{\bf Top panel}: Variance of the matter fluctuations as a function of halo mass at $z=0$. The solid orange line shows an idealised reference base from computing $\sigma(M)$ as given in Eq.~(\ref{eq:3.5}), using a continuum of Fourier modes with a large-scale cutoff set at the Horizon scale $k_{\rm min}=10^{-4}\kMpc$ and a small-scale cutoff of $k_{\rm max}=10^{2}\kMpc$. The dashed lines shows the variance computed using only the linear power spectrum, evolved to the present day that is sampled on the Fourier lattice when generating the Gaussian random field for the initial conditions, with different box size $L$ and grid size $N$. The vertical dotted lines correspond to $M_{\rm Ny}$, the mass-scale associated with haloes whose linear theory radius is given by the Nyquist scale of the initial Fourier mesh. {\bf Bottom panel}: fractional difference of the variance measures with respect to the continuum case. The line-styles are as above.}
    \label{fig:sigma_lattice}
\end{figure}

In our simulation setup, we generate a Gaussian random field of fluctuations defined by the linear matter power spectrum, which is realised on a finite Fourier lattice using a PM grid of size $N_{\rm g}=2048^3$. For our simulation box, this corresponds to a fundamental mode $k_{\rm f}=2\pi/L = 0.0126\,\kMpc$ and a Nyquist frequency $k_{\rm Ny} = \pi N_{\rm g}/L = 12.9\,\kMpc$. To assess the impact of missing variance from $k$-space discreteness, we follow and extend the method of \citet{Smithetal2003} to estimate the $\sigma(M)$ relation on the lattice as follows:
\begin{align} 
\sigma^2_{\rm lat}(M) &  = \sum_{lmn} \frac{(\Delta k)^3}{(2\pi)^3}
P(k_{lmn}) W^2(k_{lmn}R) \nn \\
& = \frac{2\pi^2}{V}
\sum_{l=-N_{\rm g}/2}^{N_{\rm g}/2}
\sum_{m=-N_{\rm g}/2}^{N_{\rm g}/2}
\sum_{n=-N_{\rm g}/2}^{N_{\rm g}/2}
\frac{\Delta^2(k_{lmn}) }{k_{lmn}^3} W^2(k_{lmn}R) \nn \\
& = \frac{16\pi^2}{V}
\sum_{l=0}^{N_{\rm g}/2}
\sum_{m=0}^{N_{\rm g}/2} 
\sum_{n=0}^{N_{\rm g}/2}  
\frac{\Delta^2(k_{lmn}) }{k_{lmn}^3} W^2(k_{lmn}R)\ , \label{eq:siglat}
\end{align}
where $k_{lmn}\equiv 2\pi/L (l^2+m^2+n^2)^{1/2}$ and $\Delta^2(k)\equiv 4\pi k^3P(k)/(2\pi)^3$.  Note that in the second line the upper and lower limit of the sums is given by the Nyquist frequency normalised by the fundamental mode. The final line uses the fact that the power spectrum and filter scale depend only on the magnitude of the $k$-vector and so the inherent symmetry means that we only need to compute the sum over one octant. In practice, we individually compute the variance of the interior $k$-points, the planes and axes, and then multiply by their corresponding weights. The others are identical, either due to symmetry, rotation or parity invariance. Also note that $\lim_{k\rightarrow0} \Delta^2(k)/k^3\rightarrow 0$, for CDM-like power spectra.  This enables the lattice computation to be sped up by an order of magnitude. Finally, we note that this would not be true if we instead used the actual realisation of the power spectrum for the initial conditions, since in this case we would only have the symmetry $\delta^*(\bk) = \delta(-\bk)$, for a real field, which means that only the upper half of $k$-space is independent. 

Figure \ref{fig:sigma_lattice} shows that the results computed using the Fourier lattice agree to high precision for $M \gtrsim M_{\rm Ny}$, where we define the Nyquist mass scale as:
\begin{align} 
M_{\rm Ny} & = \frac{4}{3}\pi \rhob \left( \frac{2\pi}{k_{\rm Ny}}\right)^3 
= \frac{4}{3}\pi \rhob \left(\frac{2\pi L}{\pi N_{\rm g}}\right)^3 
= \frac{32}{3}\pi m_{\rm p} \left(\frac{N_{\rm part}}{N_{\rm g}}\right)^3 \ ,
\end{align}
which corresponds to the range where the haloes are resolved by more than $\sim 33$ particles. Given that the haloes we considered in this work are sampled by at least 150 particles, the impact of the missing variance owing to the Fourier lattice can be considered to be negligible.

\section{Effect of binning on multiplicity function}
\label{app:bining}

\begin{figure*}
    \includegraphics[width=2.0\columnwidth]{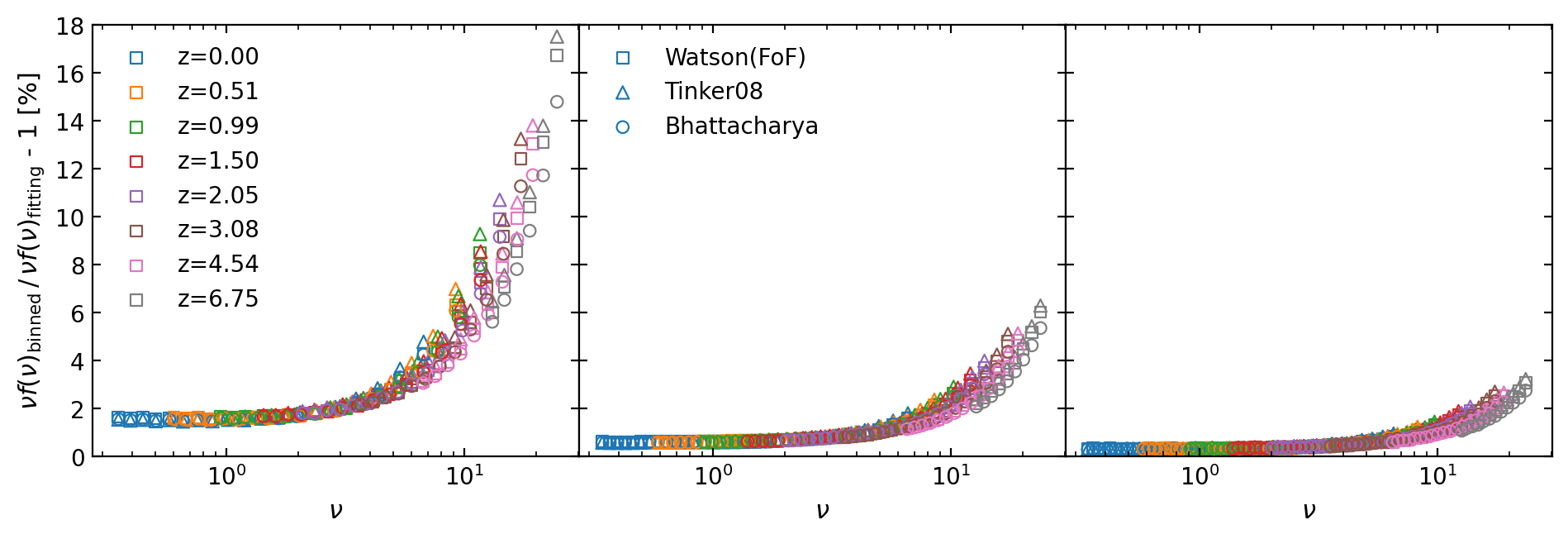}
    \caption{Relative difference of the binned $\nu f(\nu)$ to fitting
      $\nu f(\nu)$ as a function of $\nu$. The left, middle, and right
      panels are computed using, respectively, 5, 8, and 11 bins per
      decade in logarithmic scale of halo masses. The squares,
      triangles and circles denote the fitting functions of
      Watson(FoF), Tinker08, and Bhattacharya, respectively. The
      different colours of points correspond to the redshifts shown in
      the legend.}
    \label{fig:B1}
\end{figure*}

We quantified the systematic bias on the binned mass function
regarding bin width. Firstly, we assume some forms of the fitting
functions as true mass functions, which are obtained by a HMF
calculator {\tt hmf} \citep[]{Murrayetal2013}. Then we adopt our
fiducial cosmology to estimate the binned mass functions from the
fitting functions by computing the numerical integral of each bin
over its bin width. The relative differences between the binned $\nu
f(\nu)$ and the fitting $\nu f(\nu)$ are shown in Fig.~\ref{fig:B1},
which indicates that binning would cause an overestimate of $\nu
f(\nu)$. This can be seen with all 3 selected models of fitting
functions, i.e., Watson(FoF) (squares), Tinker08 (triangles), and
Bhattacharya (circles), although the extent of the overestimate could
be a little different regarding each model. The left, middle and right
panels are for a choice of 5, 8, and 11 bins per decade, respectively,
in logarithmic scale of halo masses. By comparing them, it is
understood that the bias can be theoretically reduced by increasing
the number of bins applied. However, in practice, small bin width
could cause a problem of empty bins as well as very {fluctuating} mass
function when counting haloes in numerical simulations. Thus an
optimal choice of bin width should be checked case by
case. Additionally, it can be seen that the binned mass functions tend
to be much more biased towards high redshift with the same choice of
bin width. For example, an overestimate is below $2 \%$ at $z=0$ with
8 bins per decade used, whereas it could reach a maximum of $6 \%$ at
$z=6.75$. Therefore, if one tries to measure an actual mass function
with binning method, the correction of the bias respecting bin width
should be seriously taken into account.


\section{Results of multiplicity function for individual runs of fiducial cosmology}
\label{app:nufnufiducial}

\begin{figure}
	\includegraphics[width=1.\columnwidth]{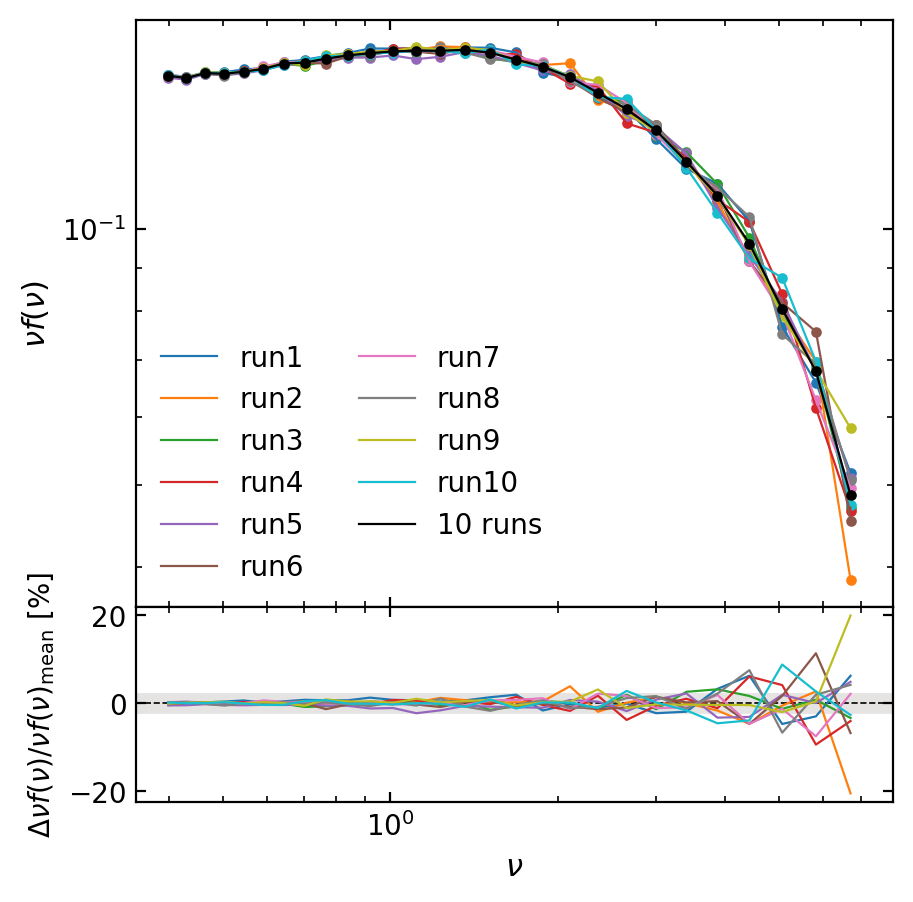}
    \caption{{\bf Top panel:} Comparison of the measured $\nu
    f(\nu)$ as a function of $\nu$ for
      10 individual runs (coloured lines) and the mean of them (black line) in the fiducial cosmology at $z=0$. {\bf Bottom panel:} Fractional difference of the data of each run relative to the mean. The grey area marks the
      relative difference of $\pm 2\%$. Lines are as before. }
    \label{fig:D}
\end{figure}

Fig.~\ref{fig:D} shows a comparison of the measured $\nu f(\nu)$ for 10 individual runs and the mean of them in the fiducial cosmology at $z=0$. Each individual realisation exhibits distinct features, particularly for $\nu \gtrsim 2$, where deviations from the mean locus gradually exceeds $2 \%$. However, when averaging over all 10 runs, the resulting $\nu f(\nu)$ locus appears significantly smoother.

\section{Results of correlation matrices for individual runs of fiducial cosmology}
\label{app:corr}

\begin{figure*}
	\includegraphics[width=1.6\columnwidth]{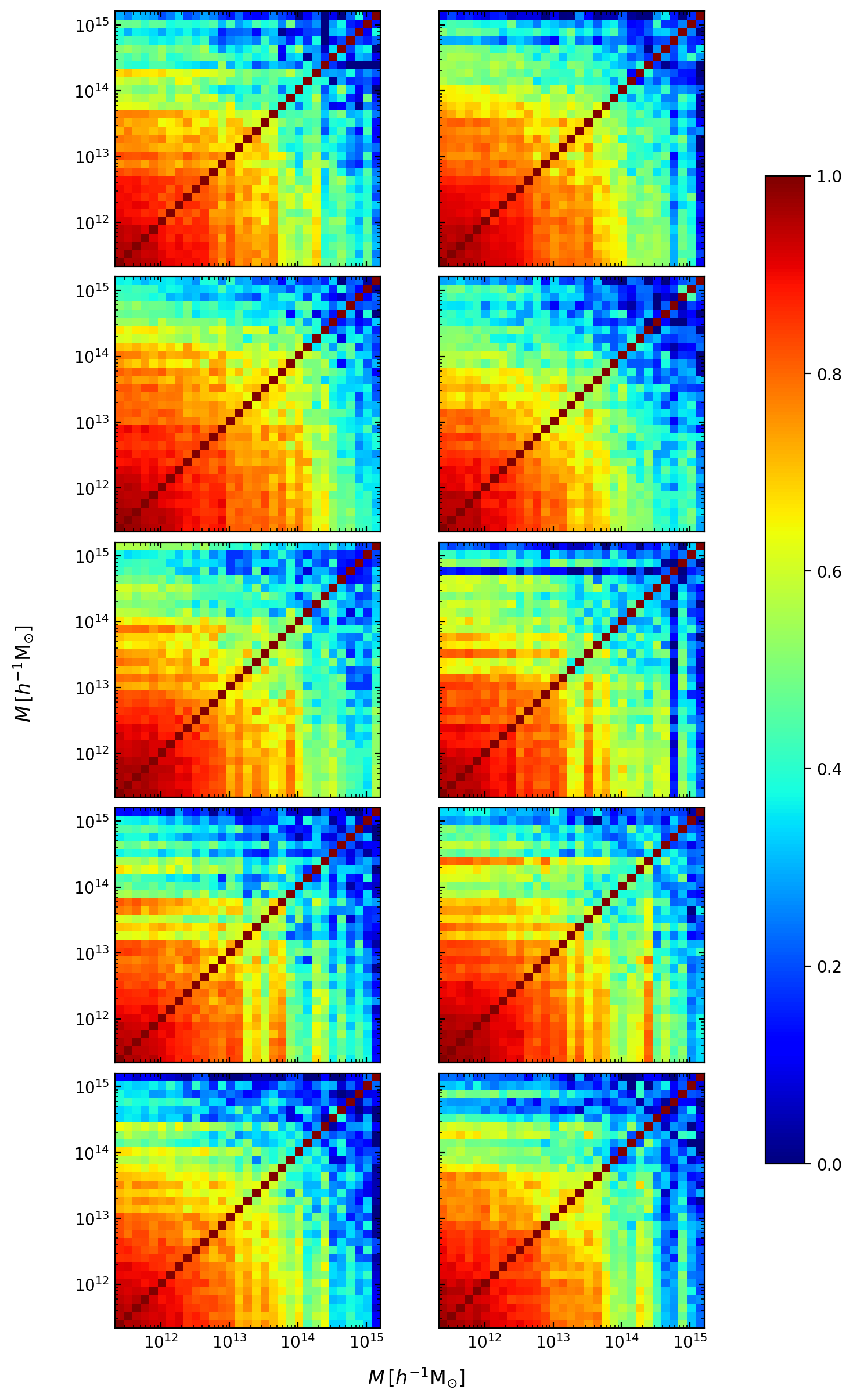}
    \caption{Correlation matrices of ${\rm d} n / {\rm d} \log{M}$ for
      10 individual runs of the fiducial cosmology at $z=0$.}
    \label{fig:C}
\end{figure*}

Fig.~\ref{fig:C} represents the results of correlation matrices of
${\rm d} n / {\rm d} \log{M}$ for each of the 10 runs of the fiducial
cosmology at $z=0$.

\section{Results of multiplicity function for higher redshifts}
\label{app:nufnuhighz}

\begin{figure*}
	\includegraphics[width=1.97\columnwidth]{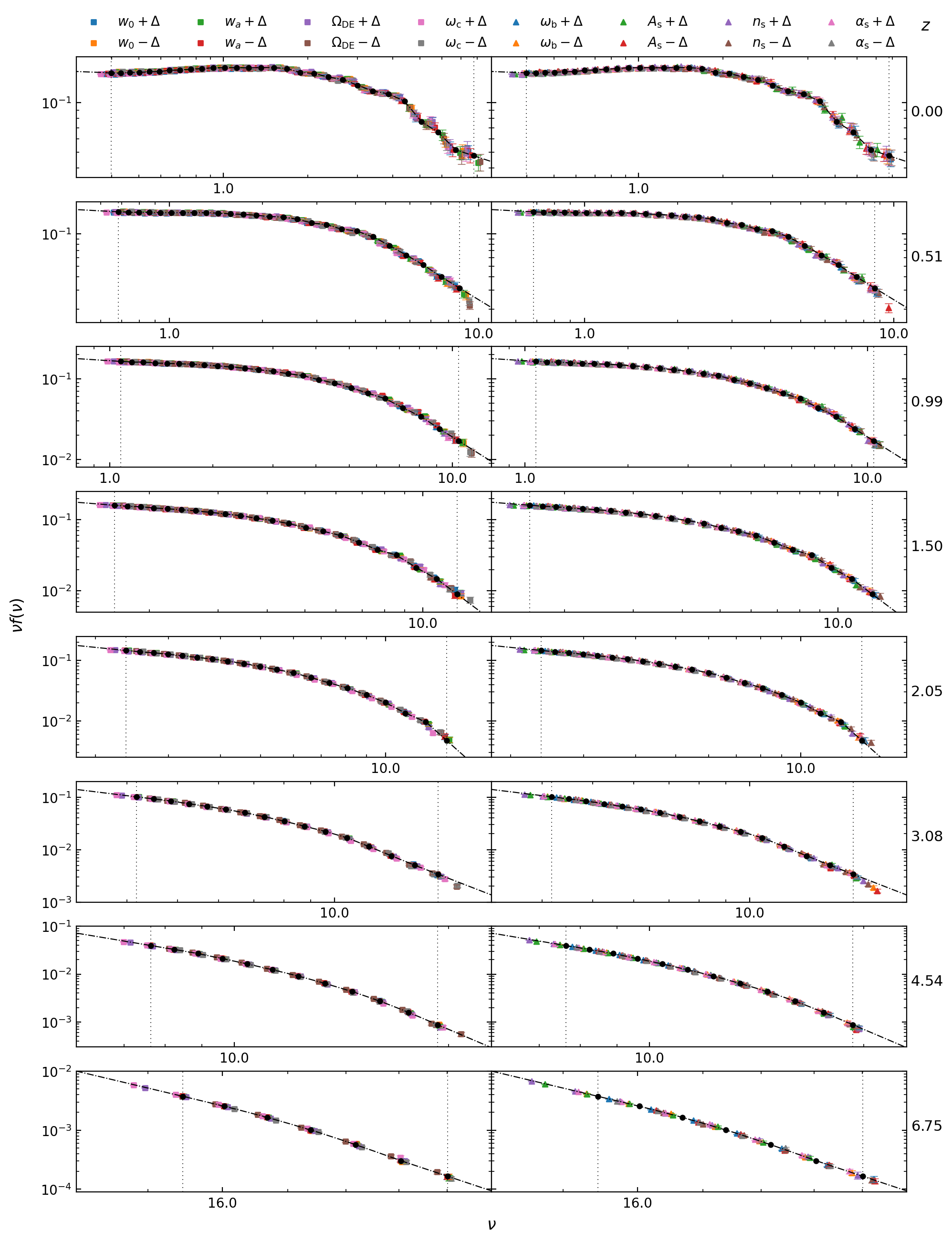}
    \caption{Comparison of $\nu f(\nu)$ as a function of $\nu$ at
      different redshifts. Each panel shows the $\nu f(\nu)$ measured
      from the fiducial run (black points) and variational runs
      (coloured squares or triangles), as well as the assumed
      universal locus (black dash-dotted line) estimated using linear
      interpolator. The two black dotted vertical lines delineate the
      most left and right black points, the locus beyond which is
      considered to be invalid. The corresponding redshifts are shown
      on the right side of each row. The measured $\nu f(\nu)$ from
      all models are very close to the loci, especially going to high
      redshift.}
    \label{fig:A1}
\end{figure*}

Fig.~\ref{fig:A1} shows the results of $\nu f(\nu)$ at $0 \leq z
\leq 6.75$. The assumed universal loci are {fluctuating} at low
redshift and tend to be smooth towards high redshift. It can be seen
that the $\nu f(\nu)$ measured from all models are very close to the
universal locus at each of the redshifts, especially going to high
redshift.


\label{lastpage}
\end{document}